\documentclass[final,finalrunningheads,a4paper]{llncs}

\usepackage[T1]{fontenc}

\usepackage{color}
\usepackage[usenames,dvipsnames]{xcolor}

\usepackage[framemethod=tikz]{mdframed}

\usepackage{multicol}
\usepackage{booktabs}
\usepackage{siunitx}
\usepackage{textcomp}

\usepackage[most]{tcolorbox}

\tcbset{myoneside/.style={
colback=LightGrey,
colframe=gray,
top=0pt,
left=2pt,
right=2pt,
bottom=-5pt,
boxsep=1pt,
halign=flush center,
nobeforeafter,
before skip=-5pt,
after skip=0pt,
boxrule=1pt
}}

\tcbset{mytwosides/.style={
colback=LightGrey,
colframe=gray,
top=0pt,
left=2pt,
right=2pt,
bottom=-5pt,
boxsep=1pt,
sidebyside,
sidebyside align=top,
sidebyside gap=8pt,
halign=flush center,
nobeforeafter,
before skip=-5pt,
after skip=0pt,
boxrule=1pt
}}

\usepackage{pst-func}
\usepackage{amssymb}
\usepackage{pifont}
\usepackage{graphicx}
\usepackage{svg}

\usepackage{environ}

\usepackage{tikz}
\usepackage{forest}
\usetikzlibrary{shadows}
\usetikzlibrary{matrix,chains,positioning,decorations.pathreplacing}
\usetikzlibrary{patterns,calc,fit,arrows}
\usetikzlibrary{shapes.geometric}
\usetikzlibrary{positioning,arrows.meta,backgrounds}

\usepackage{mathtools}
\usepackage{array,multirow}
\usepackage{algorithm}

\usepackage{mdframed}
\usepackage{arydshln}
\usepackage{etoolbox}
\usepackage{xspace} 
\usepackage{dirtytalk} 
 
\usepackage{makecell}
\usepackage{stackengine}
\setstackEOL{\cr} 

\usepackage[inline,shortlabels]{enumitem} 
\newlist{inlinelist}{enumerate*}{1}
\setlist*[inlinelist,1]{%
  label=(\roman*),
}

\usepackage{nicefrac}

\usepackage{xifthen}  
\usepackage{ifthen}

\usepackage[hidelinks]{hyperref}
\usepackage{cleveref}
\usepackage{bigints}
\hypersetup{final} 

\usepackage[final,nomargin,inline,index]{fixme} 
\fxusetheme{color}

\FXRegisterAuthor{bart}{anbart}{\color{magenta} {\underline{bart}}}
\FXRegisterAuthor{enrico}{anenrico}{\color{red} {\underline{enrico}}}

\usepackage{hhline}
\usepackage{array,multirow}
\usepackage[disable]{todonotes}

\hypersetup{
  breaklinks   = true,
  colorlinks   = true, 
  urlcolor     = blue, 
  linkcolor    = blue, 
  citecolor    = red   
}
\hypersetup{final} 

\usepackage{caption}
\usepackage{subcaption}

\DeclareCaptionType{listing}[Listing][List of Listings] 
\crefname{listing}{Listing}{Listings}  

\usepackage{xurl} 

\usepackage{listings}

\usepackage[misc]{ifsym}
\usepackage{orcidlink}

\definecolor{listingBG}{HTML}{FFFFCB}%
\definecolor{listingFrame}{HTML}{BBBB98}%
\definecolor{listingLineno}{rgb}{0.5,0.5,1.0}%

\definecolor{LightGrey}{rgb}{0.975,0.975,0.975}

\lstdefinelanguage{solidity}{
, basicstyle=\ttfamily\linespread{1.15}\normalsize\lst@ifdisplaystyle\fontsize{8}{9}\selectfont\fi
, commentstyle=\color{Gray}
, morecomment=[l]{//}
, morecomment=[s]{/*}{*/}
, escapechar=\$
, escapeinside={(*}{*)}
, classoffset=0,
, keywordstyle=\color{NavyBlue}\bfseries
, morekeywords={assert,require,if,then,else,for,break,call,delegatecall,transfer,send,approve,receive,balanceOf,case, catch,continue,do,while,emit, new, return, revert, selfdestruct, try, with, throw, switch, suicide}
, classoffset=1
, keywordstyle=\color{YellowGreen}\bfseries
, morekeywords={external, implements, import, interface, internal, library, payable, pragma, private, protected, public, pure, returns, super, using, view, immutable, memory}
, classoffset=2
, keywordstyle=\color{blue}
, morekeywords={rule,function, constructor, contract, constant, struct, address, bool, byte, bytes, bytes1, bytes2, bytes3, bytes4, bytes5, bytes6, bytes7, bytes8, bytes9, bytes10, bytes11, bytes12, bytes13, bytes14, bytes15, bytes16, bytes17, bytes18, bytes19, bytes20, bytes21, bytes22, bytes23, bytes24, bytes25, bytes26, bytes27, bytes28, bytes29, bytes30, bytes31, bytes32, enum, int, int8, int16, int24, int32, int40, int48, int56, int64, int72, int80, int88, int96, int104, int112, int120, int128, int136, int144, int152, int160, int168, int176, int184, int192, int200, int208, int216, int224, int232, int240, int248, int256, mapping, string, uint, uint8, uint16, uint24, uint32, uint40, uint48, uint56, uint64, uint72, uint80, uint88, uint96, uint104, uint112, uint120, uint128, uint136, uint144, uint152, uint160, uint168, uint176, uint184, uint192, uint200, uint208, uint216, uint224, uint232, uint240, uint248, uint256, var, void, ether, finney, szabo, wei, days, hours, minutes, seconds, weeks, years, token}
, classoffset=3
, keywordstyle=\color{Plum}\bfseries
, morekeywords={balance, block, blockhash, instanceof, coinbase, difficulty, gaslimit, number, timestamp, msg, data, gas, sender, value, sig, value, now, tx, gasprice, origin}
}

\lstdefinelanguage{cvl}{
	commentstyle=\color{Gray},
	morecomment=[l]{//},
	morecomment=[s]{/*}{*/},
	classoffset=0,
        escapechar=\$,
	morekeywords={anonymous, assembly, balance, break, call, callcode, case, catch, class, constant, continue, constructor, contract, debugger, default, delegatecall, delete, do, else, emit, event, experimental, export, external, false, finally, for, function, gas, if, implements, import, in, indexed, instanceof, interface, internal, is, length, library, log0, log1, log2, log3, log4, memory, modifier, new, payable, pragma, private, protected, public, pure, push, return, returns, revert,  selfdestruct, send, solidity, storage, struct, suicide, super, switch, then, this, throw, transfer, true, try, typeof, using, value, view, while, with, addmod, ecrecover, keccak256, mulmod, ripemd160, sha256, sha3},
	keywordstyle=\color{NavyBlue}\bfseries,
	classoffset=1,
	morekeywords={address, bool, byte, bytes, bytes1, bytes2, bytes3, bytes4, bytes5, bytes6, bytes7, bytes8, bytes9, bytes10, bytes11, bytes12, bytes13, bytes14, bytes15, bytes16, bytes17, bytes18, bytes19, bytes20, bytes21, bytes22, bytes23, bytes24, bytes25, bytes26, bytes27, bytes28, bytes29, bytes30, bytes31, bytes32, enum, int, int8, int16, int24, int32, int40, int48, int56, int64, int72, int80, int88, int96, int104, int112, int120, int128, int136, int144, int152, int160, int168, int176, int184, int192, int200, int208, int216, int224, int232, int240, int248, int256, mapping, string, uint, uint8, uint16, uint24, uint32, uint40, uint48, uint56, uint64, uint72, uint80, uint88, uint96, uint104, uint112, uint120, uint128, uint136, uint144, uint152, uint160, uint168, uint176, uint184, uint192, uint200, uint208, uint216, uint224, uint232, uint240, uint248, uint256, var, void, ether, finney, szabo, wei, days, hours, minutes, seconds, weeks, years},
	keywordstyle=\color{blue},
	classoffset=2,
	morekeywords={block, blockhash, coinbase, difficulty, gaslimit, number, timestamp, msg, data, gas, sender, sig, value, now, tx, gasprice, origin},
	keywordstyle=\color{Plum}\bfseries,
 	classoffset=3,
 	morekeywords={invariant,rule,assert,satisfy,require},
	keywordstyle=\color{red}\bfseries,
}

\usepackage{stmaryrd}

\newcommand{\ifempty}[3]{%
  \ifthenelse{\isempty{#1}}{#2}{#3}%
}

\usepackage{refcount}

\newcommand{\myfootnotetext}[1]{\footnotetext{#1\label{fn:text}%
        \edef\fnmark{\getpagerefnumber{fn:mark}}%
        \edef\fntext{\getpagerefnumber{fn:text}}%
        \ifx\fnmark\fntext\else\ClassWarning{}{footnote mark and text on different pages!}\fi}}

\newcommand{\codefont}{\fontsize{10}{10}\selectfont}
\newcommand{\code}[1]{{\tt\codefont {#1}}}

\newcommand{\contract}[1]{{\texttt{\txColor{#1}}}}
\newcommand{\txcode}[1]{{\texttt{\txColor{#1}}}}

\def\etc{\emph{etc}.\@\xspace}

\newcommand{\eg}{e.g.\@\xspace}
\newcommand{\ie}{i.e.\@\xspace}

\newcommand{\bnfdef}{\;\mbox{::=}\;}
\newcommand{\bnfmid}{\;|\;}

\newcommand{\bind}[2]{\nicefrac{#2}{#1}}
\newcommand{\setenum}[1]{\{#1\}}
\newcommand{\setcomp}[2]{\left\{{#1} \,\mid\, {#2}\right\}}

\newcommand{\irule}[2]{\dfrac{#1}{#2}}
\newcommand{\braceleft}{\code{\{}}
\newcommand{\braceright}{\code{\}}}

\newcommand{\range}[2]{{#1}..{#2}}

\newcommand{\sem}[2][]{\mbox{\ensuremath{\llbracket{#2}\rrbracket_{#1}}}}
\newcommand{\semcmd}[3][]{\mbox{\ensuremath{\langle{#2,#3}\rangle_{#1}}}}
\newcommand{\semexp}[2][]{\sem[#1]{#2}}

\def\txColor{\color{blue}}

\newcommand{\txFmt}[1]{{\txColor{\sf #1}}}

\newcommand{\tx}[2][]{\txFmt{#2}_{\txColor{#1}}}
\newcommand{\txT}[1][]{\tx[#1]{T}} 

\newcommand{\txexp}[5]{{\pmv{#1}}\,:\,{\contract{#2}}.{\txcode{#3}}({#4})\ifempty{#5}{}{\${#5}}}

\def\pmvColor{\color{red}}
\newcommand{\pmvFmt}[1]{{\pmvColor{\tt #1}}}
\newcommand{\pmv}[2][]{\pmvFmt{#2}_{\pmvColor{#1}}\xspace}
\newcommand{\pmvA}[1][]{\pmv[{#1}]{A}} 
\newcommand{\pmvB}[1][]{\pmv[{#1}]{B}}

\newcommand{\pmvM}[1][]{\pmv[{#1}]{M}}

\def\cmvColor{\color{blue}}
\newcommand{\cmvFmt}[1]{{\cmvColor{\code{#1}}}}

\newcommand{\CmvU}[1][]{\cmvFmt{\mathbb{A}}_{\cmvColor{c}}} 

\DeclareMathSymbol{:}{\mathord}{operators}{"3A}

\def\addrColor{\color{black}}
\newcommand{\addrFmt}[1]{{\addrColor{\texttt #1}}}
\newcommand{\addr}[2][]{\addrFmt{#2}_{\addrColor{#1}}\xspace}
\newcommand{\addrA}[1][]{\addr[{#1}]{X}} 
\newcommand{\addrB}[1][]{\addr[{#1}]{Y}} 

\def\sysColor{\color{Black}}

\newcommand{\sysFmt}[1]{{\sysColor{#1}}}
\newcommand{\sysS}[1][]{\mathord{\sysFmt{S}_{\sysColor{#1}}}}

\newcommand{\sysSi}[1][]{\mathord{\sysColor{\sysS'_{#1}}}}

\newcommand{\sysSS}[1][]{\mathord{\sysFmt{\Sigma}_{\sysColor{#1}}}}

\newcommand{\st}[1][]{\sigma_{#1}}
\newcommand{\sti}[1][]{\sigma'_{#1}}
\newcommand{\stii}[1][]{\sigma''_{#1}}

\newcommand{\true}{\mathit{true}}
\newcommand{\false}{\mathit{false}}
\newcommand{\truesmt}{\top}
\newcommand{\falsesmt}{\bot}
\newcommand{\Forall}{\displaystyle\mathop{\forall}}
\newcommand{\Exists}{\displaystyle\mathop{\exists}}

\newcommand{\skipE}{\texttt{skip}}
\newcommand{\reqE}[1]{\texttt{require}\ifempty{#1}{}{\,#1}}
\newcommand{\ifE}[3]{\texttt{if}~{#1}~\texttt{then}~{#2}~\texttt{else}~{#3}}
\newcommand{\walE}[1]{\ifempty{#1}{}{{#1}.}\texttt{balance}}
\newcommand{\nullE}{\texttt{null}}

\newcommand{\thisE}{\texttt{this}\xspace}
\newcommand{\senderE}{\texttt{sender}}
\newcommand{\valueE}{\texttt{value}}

\newcommand{\callE}[4]{{#1}\hspace{1pt}.\hspace{1pt}{#2}({#3})\ifempty{#4}{}{\${#4}}}

\newcommand{\nullSem}{\mathit{null}}

\newcommand{\cvc}{cvc5\xspace}

\newcommand{\kindtwo}{\mbox{Kind\hspace{2pt}2}\xspace}

\newif\ifliq
\liqtrue

\newcommand{\type}[1]{\texttt{#1}}

\newcommand{\txcontract}{\mathit{cx}\xspace}
\newcommand{\txfunction}{\mathit{fx}\xspace}
\newcommand{\xavar}{\mathit{tx.sender}\xspace}
\newcommand{\xnvar}{\mathit{tx.value}\xspace}

\newcommand{\constructor}{\code{constructor}\xspace}

\newcommand{\typeT}[1][]{\tau_{#1}}
\newcommand{\boolT}{\texttt{bool}}
\newcommand{\intT}{\texttt{int}}
\newcommand{\addressT}{\texttt{address}}
\newcommand{\methodT}{\texttt{proc}}
\newcommand{\calldataargsT}{\texttt{args}\xspace}
\newcommand{\welltyped}[1]{\vdash {#1}}
\newcommand{\hastype}[0]{\,\colon}

\newcommand{\Addresses}{\mathbf{Addr}\xspace}
\newcommand{\Contracts}{\mathbf{Contr}\xspace}
\newcommand{\Methods}{\mathbf{Proc}\xspace}

\newcommand{\blocknumSMT}[1][]{\ensuremath{block\_num}}
\newcommand{\balanceSMT}[1][]{\ensuremath{balance\ifempty{#1}{}{(#1)}}}
\newcommand{\Balances}[1][]{\ensuremath{Balances}}

\newcommand{\constructed}{\ensuremath{constructed}\xspace}
\newcommand{\lastRevert}{\ensuremath{revert}\xspace}

\newcommand{\addressZero}{\text{null}}

\newcommand{\Init}{\ensuremath{\mathit{Init}}\xspace}
\newcommand{\Trans}{\ensuremath{\mathit{Trans}}\xspace}
\newcommand{\StateVars}{\ensuremath{\mathbf{SVars}}\xspace}
\newcommand{\TransVars}{\ensuremath{\mathbf{TVars}}\xspace}
\newcommand{\NextVars}{\ensuremath{\mathbf{nxSVars}}\xspace}
\newcommand{\FieldsVars}{\ensuremath{\StateVars_F}\xspace}
\newcommand{\ParamVars}{\ensuremath{\TransVars_P}\xspace}
\newcommand{\EqParams}{\ensuremath{\mathit{EqParams}}}

\newcommand{\defvalue}{\ensuremath{\mathrm{default\_value}}\xspace}
\newcommand{\encS}{\Phi}
\newcommand{\encP}{\encS}

\newcommand{\typeV}{\ensuremath{\mathrm{type}}\xspace}

\newcommand{\defas}{\mathrel{\mathop:}=} 
\newcommand{\ITE}{\mathrm{ITE}}
\newcommand{\switch}{\mathrm{switch}\xspace}
\newcommand{\mycase}{\mathrm{case}\xspace}

\newcommand{\Tree}{\ensuremath{T}\xspace}
\newcommand{\node}{\ensuremath{V}\xspace}
\newcommand{\nodei}{\ensuremath{{V'}}\xspace}
\newcommand{\edge}{\ensuremath{E}\xspace}

\newcommand{\procF}{\txcode{f}}

\newcommand{\procG}{\txcode{g}}

\newcommand{\txout}[3]{{#1}.\texttt{transfer}({#2\ifempty{#3}{}{:#3}})}

\newcommand{\lastrevL}[0]{\texttt{reverted}}

\newcommand{\oldL}[1]{\texttt{old}({#1})}
\newcommand{\txnext}[1]{\langle{#1}\rangle}
\newcommand{\alltxnext}[1]{\left[{#1}\right]}
\newcommand{\txL}[5]{{#1}\,:\,{#2}.{#3}({#4})\ifempty{#5}{}{\${#5}}}

\newcommand{\Act}{Act}

\newcommand{\hd}[1]{\mathop{hd}({#1})}
\newcommand{\tl}[1]{\mathop{tl}({#1})}

\def\validColor{\color{ForestGreen}}
\def\invalidColor{\color{red}}

\newcommand{\valid}[1]{{\validColor \checkmark} ($k=#1$)}
\newcommand{\invalid}[1]{{\invalidColor \xmark} ($#1$ steps)}

\newcommand{\statusValid}{{\validColor\checkmark}}
\newcommand{\statusInvalid}{{\invalidColor\xmark}}

\newcommand{\timelimit}{1000}

\newcommand{\myparagraph}[1]{\ \\\textbf{#1}.}
\newcommand{\myparagraphB}[1]{\ \\\emph{\normalsize{#1}}.}

\newcommand{\ourlogic}{CHML\xspace}

\newcommand{\ourtool}{KindHML\xspace} 

\newcommand{\ourtoolgithub}{https://github.com/SmartContractsVerification/KindHML}
\newcommand{\exCex}{https://github.com/SmartContractsVerification/KindHML/blob/main/results/cex_example.txt}

\newtoggle{arxiv}
\toggletrue{arxiv}

\begin{document}

\hyphenation{block-chain}
\hyphenation{Block-chain}
\hyphenation{block-chains}

\begin{frontmatter}

\title{\ourtool: formal verification of smart contracts based on Hennessy-Milner logic}

\newcommand{\emailaddress}{enrico.lipparini@unica.it}

\author{Massimo Bartoletti\orcidlink{0000-0003-3796-9774}\inst{1}
\and Angelo Ferrando\orcidlink{0000-0002-8711-4670}\inst{2} 
\and Enrico Lipparini\orcidlink{0009-0009-0428-4403}\href{mailto:\emailaddress}{$^{\textrm{(\Letter)}}$}\inst{1} 
\and Vadim Malvone\orcidlink{0000-0001-6138-4229}\inst{4}}

\institute{
Universit\`a degli Studi di Cagliari, Cagliari, Italy
\and 
Universit\`a degli Studi di Modena e Reggio Emilia, Modena, Italy
\and 
T\'el\'ecom Paris, Institut Polytechnique de Paris, France}

\maketitle

\renewcommand{\thelstlisting}{\arabic{lstlisting}}




\begin{abstract}
Smart contracts deployed on blockchains such as Ethereum routinely manage large amounts of assets, making their security critical. 
Empirical studies show that real-world attacks often exploit flaws in the business logic of contracts that unfold across multiple transactions, such as liquidity or front-running attacks. 
Detecting these attacks requires reasoning about expressive temporal properties beyond the capabilities of existing analysis tools.
In this paper, we present an automated approach to the formal verification of smart contracts, enabling the specification and verification of complex temporal properties. 
\iftoggle{arxiv}{
	Our approach provides a fully automated encoding into Lustre --- the specification language supported by the Kind 2 model checker --- of an expressive subset of Solidity contracts and temporal specifications based on first-order Hennessy-Milner Logic.
	This encoding allows us to leverage \kindtwo to determine whether the contract respects the specification or not.
}
{
	We provide a fully automated encoding of an expressive subset of Solidity and of first-order Hennessy-Milner Logic specifications into Lustre.
	This allows us to leverage \kindtwo for model checking, producing counterexamples when properties are violated and proof certificates when they hold. \enriconote{parlare di "controesempi" è potenzialmente problematico in un contesto con quantificatori (il tool ritorna solo lo stato in cui la proprietà è falsa, ma non le eventuali istanziazioni dei quantificatori necessarie); per i proof certificates sarebbe vero ma a livello pratico mi sembra che ci sia qualche problema (devo indagare meglio)	}
}
\iftoggle{arxiv}
{%
	We implement our approach in a toolchain that integrates the translation and verification steps, and we evaluate its effectiveness and performance on a benchmark of smart contracts and temporal properties capturing complex attack scenarios.
}
{
	We implement our approach in a toolchain integrating translation and verification, and evaluate it on a benchmark of smart contracts and temporal properties capturing complex attack scenarios.
}
Our results show that the \iftoggle{arxiv}{proposed approach can effectively}{approach can} verify non-trivial temporal properties of smart contracts and detect violations that are beyond the reach of existing analysis tools.
\end{abstract}

\end{frontmatter}

\section{Introduction}


In recent years we have seen a steady rise of smart contracts that implement financial ecosystems on top of public blockchains, and control today tens of billions of dollars worth of crypto-assets. 
The peculiarities of the setting (\ie, the absence of intermediaries, the immutability of code after deployment, some problematic design choices in smart contract languages) make smart contracts an appealing target for attackers, as bugs might be exploited to steal crypto-assets or just cause disruption.
This is witnessed by a long history of attacks, which overall caused losses exceeding 6 billions of dollars~\cite{Chaliasos24icse}.

Several vulnerability detection tools for smart contracts have been developed in recent years. 
For Ethereum --- largely the main smart contract platform --- dozens are available~\cite{Tolmach22csur,Kushwaha22access,Zhang23icse}, 
spanning techniques from static analysis and symbolic execution to fuzzing, and, recently, machine-learning~\cite{RESSI2025100390} and LLMs~\cite{LLM-SmartAudit,SmartGuard}.
%
However, their effectiveness in preventing real-world attacks remains debated. 
Despite the availability of multiple vulnerability detection tools, attacks against smart contracts continue to proliferate, 
exploiting subtle flaws in the contracts' business logic rather than well-known vulnerability patterns.
Empirical studies confirm this trend. 
For instance, the analysis in~\cite{Chaliasos24icse} shows that existing vulnerability detection tools could have prevented only about 8\% of $\$2.3$ billion in total losses due to attacks, and only when these attacks exploited well-known vulnerabilities such as reentrancy.
The problem is that many high-impact incidents stem instead from \emph{logic errors} in contract implementations --- \ie, flaws in   protocol logic or economic rules. 
Such errors typically fall outside the scope of vulnerability detection tools, which are mainly designed to identify specific classes of bugs (\eg,  overflows and reentrancy).
In contrast, formal verification offers a principled defense against these attacks, as it enables developers to verify that smart contracts satisfy a specification capturing their intended behaviour.

For Solidity, the most widely adopted smart contract language, a few formal verification tools have been developed that can verify (or refute) user-defined properties concerning the business logic of a contract. 
Notable tools include SolCMC~\cite{AltS22cav}, shipped with the Solidity compiler, and the Certora Prover~\cite{certora}, a leading tool in the smart contract auditing industry. 
However, a key limitation of these tools lies in the expressiveness of the properties they support~\cite{BFMPS24fmbc,BCL25fmbc}.
For example, several real-world attacks exploit \emph{liquidity} weaknesses in smart contracts, whereby an attacker drives the contract into a state in which a legitimate user can no longer perform an otherwise permitted action, such as withdrawing tokens. 
Such vulnerabilities have been exploited in practice to steal or permanently freeze crypto-assets~\cite{Alois17parity}.
Another widespread type of attack is \emph{front-running}, where the adversary exploits the transaction sequencing mechanism in order to obtain a profit at the expense of legit users~\cite{Eskandari19sok,Torres21frontrunner}. 
Both liquidity and front-running attacks involve behaviours that span multiple transactions and depend on the evolution of the contract state over time. Capturing these vulnerabilities therefore requires reasoning about temporal properties --- for instance, that certain actions remain eventually enabled for honest users, or that an adversary who front-runs a user's transaction cannot break some desired invariant. 
Such properties go beyond those typically supported by existing smart contract verification tools such as SolCMC and the Certora Prover.


\paragraph{Contributions}

We propose \ourtool, a tool that verifies complex temporal properties of smart contracts. 
\ourtool takes as input a contract, written in a purified version of Solidity, and a set of user-defined properties. 
Such properties are expressed in a novel specification language (\ourlogic), which extends first-order Hennessy-Milner logic with domain-specific constructs for the smart contracts setting.
In particular, \ourlogic can express liquidity and front-running properties that are out of the scope of what can be expressed in existing verification tools for smart contracts, including industrial tools such as SolCMC and the Certora Prover, as well as academic tools.
We experiment \ourtool on a benchmark of smart contracts, showing that it can effectively verify complex properties.
\ourtool provides developers with useful feedback, by detecting logical errors that would otherwise remain unnoticed.
In particular, when \ourtool detects a property violation, it produces a concrete execution trace that leads the contract to a state from which the desired asset exchange is unrealisable. 
To the best of our knowledge, \ourtool is the first tool capable of verifying front-running properties of smart contracts.

Summing up, the main contributions of the paper include:
\begin{itemize}

\item \ourlogic, a novel temporal logic that can express complex temporal properties of smart contracts (encompassing, \eg, liquidity and front-running);

\item a fully automated toolchain that encodes \ourlogic properties and an expressive fragment of Solidity into Lustre~\cite{Lustre}, and verifies the resulting specification using \kindtwo~\cite{kind2};

\item an empirical evaluation of the effectiveness 
of the proposed approach.

\end{itemize}

Our toolchain and experimental data are publicly available on \href{\ourtoolgithub}{github}.

\iftoggle{arxiv}{
	\paragraph{Structure.} 
	The paper is organized as follows.
	In \Cref{sec:contracts} we describe the system model.
	In \Cref{sec:examples} we introduce some use cases, that we use to present \ourlogic (\Cref{sec:logic}) and the encoding of CHML to first-order logic \Cref{sec:encoding}.
	Implementation details and  experimental evaluation  are in \Cref{sec:evaluation}.
	Related work is discussed in Section~\ref{sec:related}.
	Finally, in Section~\ref{sec:conclusions}
	we draw some conclusions and discuss future work.
}
{
}

\section{System model}
\label{sec:contracts}

We consider a system model inspired by stateful account-based platforms such as Ethereum. 
In particular, we formalize a fragment of Solidity, the main contract language in the Ethereum platform.
%
For simplicity, we describe here an untyped version of the language, where type declarations are omitted, and booleans are encoded as integers.  
The concrete language we have implemented in our tool is typed, and it features a type system to rule out type mismatches.

\iftoggle{arxiv}{\subsection{Background}
\label{sec:background}

From the perspective of smart contract programming, a blockchain can be viewed as an asset-exchange state machine. 
The global state records the assets owned by each \emph{account}. 
User-created \emph{transactions} trigger state transitions that may transfer assets among accounts.

In Solidity, accounts are of two kinds: externally owned accounts (EOAs) and contract accounts. EOAs are controlled by users through cryptographic keys and can initiate transactions, while contract accounts are controlled by the code of a deployed contract and can execute transactions only in response to \emph{calls} from other accounts.
Such calls resemble method invocations in object-oriented languages, with contracts playing the role of objects whose methods can be invoked by other contracts and EOAs.
The global state of the blockchain can thus be seen as a mapping that assigns to each EOA its balance of native assets (\eg, ETH in Ethereum), and to each contract account both a balance and a \emph{storage}.
The storage contains the persistent variables and data structures (\eg, key-value maps) that overall define the contract state.
Transactions can update the balance and the storage of the invoked accounts, as well as the balance of the transaction sender.

Despite this resemblance, smart contracts differ from objects in several key aspects. 
In particular, smart contracts execute in a public, permissionless environment in which transaction ordering is determined through a consensus protocol among mutually untrusted (possibly, adversarial) nodes. 
Moreover, once deployed on the blockchain, the code of a smart contract cannot be modified. 
As a consequence, bugs discovered after deployment cannot be fixed, making it critical to verify the security of contracts before they are deployed.}{}

\iftoggle{arxiv}{%
\subsection{Syntax of the Solidity fragment}
\label{app:contracts:syntax}
}

\iftoggle{arxiv}{}
{%
\paragraph{Contracts.}
\label{sec:contracts:syntax}
}
\iftoggle{arxiv}{
	\begin{figure}[b!]
	\input{fig_syntax}
	\end{figure}
}
{}
A contract can be seen as a public API through which users exchange \emph{tokens}.
For simplicity, we assume a single token type --- typically, the native cryptocurrency provided by the underlying blockchain (\eg, ETH for Ethereum).
Users interact with a contract via \emph{addresses}, which serve as pseudonymous identifiers corresponding to their public keys.
We use colors to discriminate between different types of addresses: 
user addresses $\pmvA,\pmvB,\ldots$ are in red,
contract addresses $\contract{C}, \contract{D}$ in blue,
and arbitrary (user or contract) addresses $\addrA, \addrB, \ldots$ in black.

We specify a contract $\contract{C}$ as a finite set of procedure declarations of the form:
\(
\procF(x_1,\ldots,x_k) \braceleft s \braceright
\)
where $\procF$ is the procedure name,
$x_1,\ldots,x_k$ is the sequence of formal parameters,
and $s$ is the procedure body
(see \Cref{fig:contracts:syntax}).
We assume that all the procedures in a contract have distinct names, and a special procedure, called \txcode{constructor}, which must be called (only once) before any other call.
Statements and expressions extend those of a standard loop-free imperative language with a few domain-specific constructs:
\begin{itemize}

\item $\txout{\addrA}{n}{}$ transfers $n$ token units from the contract to address $\addrA$;

\item $\walE{\addrA}$ is a special key, associated to the storage of address $\addrA$, representing the amount of tokens owned by that address.
Unlike other keys, the balance cannot be updated through assignments;
rather, its updates are handled automatically by the contract semantics;

\item $\reqE{e}$ aborts the execution of a procedure, rolling-back its effects, if the boolean condition $e$ is false; otherwise, it continues the execution.

\end{itemize}

\iftoggle{arxiv}{%
We formalize the syntax of our Solidity fragment in~\Cref{fig:contracts:syntax}.
Note that \ourtool includes a type system that rules out ill-typed contracts.
}
{}

\iftoggle{arxiv}
{%
\subsection{Semantics of the Solidity fragment}
\label{app:contracts:semantics}
}
{%
\paragraph{Transactions.}
}

Users interact with contracts by sending \emph{transactions}, which trigger calls to the contract API.
They have the form   
\(
\txexp{A}{C}{f}{v_1,\cdots,v_k}{n}
\)
where: 
\begin{itemize}

\item $\pmvA$ is the address of the user who signed the transaction. This address can be accessed within a contract through the expression $\senderE$;  

\item $\contract{C}$ is the called contract (identified by its address);

\item $\txcode{f}$ is the called procedure, and $v_1,\cdots,v_k$ are the actual parameters;

\item $n$ is the number of tokens transferred from the sender to the contract along with the call. This value can be accessed within a contract through the expression $\valueE$.
We omit zero values for brevity.
\end{itemize}

We model the interactions between users and contracts as a transition system between \emph{blockchain states} $\st, \sti, \ldots$
A blockchain state $\st$ maps each address $\addrA$ to a \emph{storage} $\st(\addrA)$. 
For contract addresses, the storage is a key-value map that associates a value to each field (including the special field $\walE{}$).
For user addresses, the storage is a special case in which the only valid key is $\walE{}$.

Executing a transaction $\txexp{A}{C}{f}{v_1,\cdots,v_k}{n}$ in state $\st$ amounts to the following steps:
\begin{inlinelist}

\item check that the sender has sufficient balance to transfer $n$ tokens to the contract;

\item check that the procedure $\procF$ with the expected signature belongs to the contract $\contract{C}$, and retrieve its statement $s$; 

\item define a state $\stii$ that accounts for the transfer of $n$ tokens from the sender $\pmvA$ to the contract $\contract{C}$;

\item map the formal to the actual parameters of the called procedure, also taking the special variables $\senderE$, $\valueE$, and $\thisE$ into accoiunt; 

\item evaluate the statement $s$ in state $\stii$, producing a new state $\sti$.
If some $\reqE{}$ commands in the statement $s$ fail,
then the evaluation of $s$ gives $\bot$, representing an execution error.
This makes the premise false, and so the transition does not happen.

\end{inlinelist}

\iftoggle{arxiv}{}
{Because of space constraints, we relegate the formalization of the semantics of our Solidity fragment to~\Cref{app:contracts}.}
\iftoggle{arxiv}{
\iftoggle{arxiv}{}
{
	\begin{figure}[b!]
		\input{fig_syntax}
	\end{figure}
}

\iftoggle{arxiv}{}
{\section{Formalization of the Solidity fragment}
\label{app:contracts}

In this~\namecref{app:contracts} we provide a minimal background on smart contracts and on the Solidity language (\Cref{sec:background}), and then formalize the syntax and semantics  (\Cref{app:contracts:syntax,app:contracts:semantics}) of our Solidity fragment.

\subsection{Background}
\label{sec:background}

From the perspective of smart contract programming, a blockchain can be viewed as an asset-exchange state machine. 
The global state records the assets owned by each \emph{account}. 
User-created \emph{transactions} trigger state transitions that may transfer assets among accounts.

In Solidity, accounts are of two kinds: externally owned accounts (EOAs) and contract accounts. EOAs are controlled by users through cryptographic keys and can initiate transactions, while contract accounts are controlled by the code of a deployed contract and can execute transactions only in response to \emph{calls} from other accounts.
Such calls resemble method invocations in object-oriented languages, with contracts playing the role of objects whose methods can be invoked by other contracts and EOAs.
The global state of the blockchain can thus be seen as a mapping that assigns to each EOA its balance of native assets (\eg, ETH in Ethereum), and to each contract account both a balance and a \emph{storage}.
The storage contains the persistent variables and data structures (\eg, key-value maps) that overall define the contract state.
Transactions can update the balance and the storage of the invoked accounts, as well as the balance of the transaction sender.

Despite this resemblance, smart contracts differ from objects in several key aspects. 
In particular, smart contracts execute in a public, permissionless environment in which transaction ordering is determined through a consensus protocol among mutually untrusted (possibly, adversarial) nodes. 
Moreover, once deployed on the blockchain, the code of a smart contract cannot be modified. 
As a consequence, bugs discovered after deployment cannot be fixed, making it critical to verify the security of contracts before they are deployed.
}

\iftoggle{arxiv}{}
{%
\subsection{Syntax of the Solidity fragment}
\label{app:contracts:syntax}

We formalize the syntax of our Solidity fragment in~\Cref{fig:contracts:syntax}.
Note that \ourtool includes a type system that rules out ill-typed contracts. 
}

\begin{figure}[b!]
\input{fig_semanticsExpr}
\end{figure}

\begin{figure}[h]
\input{fig_semanticsSt}
\end{figure}

\iftoggle{arxiv}{}
{\subsection{Semantics of the Solidity fragment}
\label{app:contracts:semantics}}

We use the standard notation $\setenum{\bind{x}{v}}$ to represent a partial function mapping $x$ to $v$.
This can be composed with other partial functions as usual:
\eg, $\rho\setenum{\bind{x}{v}}$ is the function $\rho'$ such that
$\rho'(x) = v$ and $\rho'(y) = \rho(y)$ for $y \neq x$.
We use the syntactic sugar $\st(\addrA.x)$ to denote the value of key $x$ at address $\addrA$, 
and $\st \setenum{\bind{\addrA.x}{v}}$ to denote an update of key $x$ at address $\addrA$ in storage $\st$.
More formally, we define:
\[
\st(\addrA.x) = \st(\addrA)(x) 
\qquad
\st\setenum{\bind{\addrA.x}{v}} 
= 
\st\setenum{\bind{\addrA}{f}}
\;\;
\text{ where }
f = \st(\addrA)\setenum{\bind{x}{v}} 
\]

State transitions between blockchain states are labelled with the transactions fired by users.
The transition relation $\xrightarrow{\;\;\;}$ is specified by the following rule:
\[
    \irule
    {\begin{array}{c}
    \st(\walE{\pmvA}) \geq n
    \qquad
    \procF(x_1,\ldots,x_k) \braceleft s \braceright \in \contract{C}
    \\[4pt]
    \stii = \st
    \setenum{\bind{\walE{\pmvA}}{\st(\walE{\pmvA}) - n}}
    \setenum{\bind{\walE{\contract{C}}}{\st(\walE{\contract{C}}) + n}}
    \\[4pt]
    \rho = \setenum{\bind{x_1}{v_1} \cdots \bind{x_k}{v_k}}
           \setenum{\bind{\senderE}{\pmvA}}
           \setenum{\bind{\valueE}{n}}
           \setenum{\bind{\thisE}{\contract{C}}}
    \qquad
    \semcmd[\rho]{s}{\stii} \Rightarrow \sti
    \end{array}}
    {\st \xrightarrow{\; \txexp{A}{C}{f}{v_1,\cdots,v_k}{n} \;} \sti}
\]

We say that a transaction $\txT$ is \emph{valid} in a blockchain state $\st$ when there exists some $\sti$ such that $\st \xrightarrow{\txT} \sti$.
We write $\st \;\;\not\!\!\xrightarrow{\txT}$ when $\txT$ is \emph{not} valid in $\st$.

We now comment the five premises of the rule:
\begin{itemize}

\item The premise $\st(\walE{\pmvA}) \geq n$ checks that the sender has sufficient balance to transfer $n$ tokens to the contract.

\item The second premise checks that the procedure $\procF$ with the expected signature belongs to the contract $\contract{C}$, and retrieves its statement $s$. 

\item The second line defines a state $\stii$ that accounts for the transfer of $n$ tokens from the sender $\pmvA$ to the contract $\contract{C}$.

\item In the third line, we define a map $\rho$ from the formal to the actual parameters of the called procedure.
The map also includes bindings for $\senderE$ and $\valueE$, and for the keyword $\thisE$, which is associated to the address of the called contract.
We will use this map when giving semantics to expressions and statements. 

\item Finally, the premise
$\semcmd[\rho]{s}{\stii} \Rightarrow \sti$
evaluates the statement $s$ in state $\stii$
(\ie, the state that results from $\st$ after updating the caller and contract's balances)
producing a new state $\sti$.
If some $\reqE{}$ commands in the statement $s$ fail,
then the evaluation of $s$ gives $\bot$, representing an execution error.
This makes the premise false, and so the transition does not happen.

\end{itemize}


The semantics of expressions is defined, in a big-step style, by the partial function $\semexp[\rho,\st]{\cdot}$ in~\Cref{fig:contracts:sem-exp}.
The rules that inductively define this function are mostly standard.
We just note that the evaluation of an ill-typed expression,
such as $2 + \pmvA$ or $\contract{C}.x$ when $x$ is not a field of $\contract{C}$, is undefined.

The semantics of statements is defined, also in a big-step style, by the relation $\semcmd[\rho]{\cdot}{\st}$ in~\Cref{fig:contracts:sem-cmd}.
When this function is defined, it gives the new storage state $\sti$
obtained after evaluating the statement.
Note that the relation $\Rightarrow$ is deterministic: 
therefore, also the state transition relation $\xrightarrow{}$ is deterministic.

We say that a blockchain state is \emph{initial} when all contracts have zero balance, their integer fields are set to 0, and their address fields are set to $\nullSem$.

}

\iftoggle{arxiv}{\section{Use cases}
\label{sec:examples}

In this~\namecref{sec:examples} we describe the use cases that we will use to validate our approach.

\subsection{Bank contract}
\label{sec:bank}
\iftoggle{arxiv}{
	\begin{figure}[h!]
		\small
		\input{lst_bank}
		\vspace{-15pt}
	\end{figure}
}
{
}

Consider the simple bank contract in~\Cref{lst:bank}.
Its storage consists of a single key-value map \code{credits}, which records the number of ETH units associated to each bank's client.
Its public interface consists of two functions, \code{deposit} and \code{withdraw}, which can be called by any account at any time.

When an account invokes the function \code{deposit}, two effects occur:
\begin{inlinelist}
\item an amount of ETH specified by the caller is implicitly transferred from the caller's balance to the contract's.
In the code, the read-only variables \code{msg.value} and \code{msg.sender} denote, respectively, the amount and the caller;
\item the caller's credit recorded in the contract storage is increased by the same amount.
\end{inlinelist}
The \code{withdraw} function has a dual effect:
\begin{inlinelist}
\item decrease the number of credits of the caller;
\item transfer \code{amount} ETH to the caller. 
\end{inlinelist}
Note that if some of the implicitly required conditions are not satisfied (\eg, the \code{amount} parameter of \code{withdraw} exceeds the caller's credit), the transaction is \emph{reverted}, meaning that all effects performed during its execution are rolled back.

Despite its simplicity, the behaviour of this contract exhibits non-trivial properties, such as:
\begin{itemize}

\item any user with a positive credit can always extract ETH;

\item the effect of two consecutive calls to \code{deposit} (resp., \code{withdraw}) by the same account is equivalent to a single call to the same function;

\item the effect of a call to \code{deposit} can be reversed by executing a suitable subsequent call;

\item each call affects the credits of exactly one account;

\item front-running a user's deposit with someone else's transaction has no effect on the user's credits.

\end{itemize}

\iftoggle{arxiv}{
}
{
	\begin{figure}[t]
		\small
		\input{lst_bank}
		\vspace{-15pt}
	\end{figure}
}

Reasoning about such properties requires a tool capable of handling both universal and existential quantification, as well as predicating about transactions fired across different execution paths.
To the best of our knowledge, these features, in the level of generality required to precisely capture the properties above, are not supported by existing verification tools.

\subsection{Bet contract}
\label{sec:bet}

\iftoggle{arxiv}{
	\begin{figure}[h!]
		\input{lst_bet}
	\end{figure}
}
{
	\begin{figure}[t]
	\input{lst_bet}
	\end{figure}
}
Consider a simple bet contract with the following behaviour (see~\Cref{lst:bet}).
The constructor sets the \code{oracle} and the initial bet, \ie the amount of tokens transferred upon deployment. 
The \code{join} procedure allows anyone to join the bet, by depositing into the contract an amount of tokens equal to the initial bet.
The \code{win} procedure allows the previously joined player to withdraw the whole pot, provided that \code{rate} is greater than 100.
Finally, the \code{set} procedure allows the oracle to set the rate.


\iftoggle{arxiv}{
	\begin{figure}[b!]
\begin{lstlisting}[
		,language=solidity
		,caption={A Vault contract.}
		,label={lst:vault},
		,frame=single]
contract Vault {
  address owner; address recovery;
  uint wait_time; uint req_time;
  address receiver; int amount;
  int state; // 0 = IDLE, 1 = REQ
  
  constructor (address r,int w) payable {
    require(msg.sender!=r && w>=0);
    owner = msg.sender; recovery = r; 
    wait_time = w; state = 0 // IDLE
  }
  
  function withdraw(address rcv, int amt) {
    require(state==0); state = 1; // IDLE -> REQ
    require(amt<=balance && msg.sender==owner);
    receiver = rcv; amount = amt; 
    req_time = block.number
  }
  
  function finalize() {
    require(state==1); state = 0; // REQ -> IDLE
    require(block.number>=req_time+wait_time);
    require(msg.sender==owner);	
    receiver.transfer(amount)
  }
  
  function cancel() {
    require(msg.sender==recovery);
    require(state==1); state = 0 // REQ -> IDLE
  }
}
\end{lstlisting}
	\end{figure}
}{
}

Consider an initial state $\st[0]$ where users $\pmvA$ and $\pmvB$ have 10 tokens each.
We have the following computation, where as syntactic sugar we just write $x$ to access a field $x$ within $\contract{C}$, instead of writing $\contract{C}.x$:
\begin{align*}
    \st[0]
    & \xrightarrow{\txexp{A}{Bet}{constructor}{\pmv{M},1}{10}}
    \st[1]
    \quad\text{where}
    \\
    & \qquad \st[1] = 
    \st[0]\setenum{
    \bind{\walE{\pmvA}}{0},
    \bind{\walE{\contract{Bet}}}{10},
    \bind{\contract{Bet}.\code{oracle}}{\pmvM},
    \bind{\contract{Bet}.\code{rate}}{1}
    }
    \\
    & \xrightarrow{\txexp{B}{Bet}{join}{}{10}}
    \st[2] = 
    \st[1]\setenum{
    \bind{\walE{\pmvB}}{0},
    \bind{\walE{\contract{Bet}}}{20},
    \bind{\contract{Bet}.\code{player}}{\pmvB}
    }
    \\
    & \xrightarrow{\txexp{M}{Bet}{set}{150}{}}
    \st[3] = 
    \st[2]\setenum{
    \bind{\contract{Bet}.\code{rate}}{150}
    }
    \\
    & \xrightarrow{\txexp{B}{Bet}{win}{}{}}
    \st[4] =
    \st[3]\setenum{
    \bind{\pmvB.\code{balance}}{20},
    \bind{\walE{\contract{Bet}}}{0}
    }
\end{align*}

\noindent
At the end of the execution, $\pmvB$ has withdrawn the entire pot from the contract.

\subsection{Vault contract}
\label{sec:vault}

\iftoggle{arxiv}{
}{
	\begin{figure}[t]
\begin{lstlisting}[
	,language=solidity
	,caption={A Vault contract.}
	,label={lst:vault},
	,frame=single]
contract Vault {
  address owner; address recovery;
  uint wait_time; uint req_time;
  address receiver; int amount;
  int state; // 0 = IDLE, 1 = REQ
  
  constructor (address r,int w) payable {
    require(msg.sender!=r && w>=0);
    owner = msg.sender; recovery = r; 
    wait_time = w; state = 0 // IDLE
  }
  
  function withdraw(address rcv, int amt) {
    require(state==0); state = 1; // IDLE -> REQ
    require(amt<=balance && msg.sender==owner);
    receiver = rcv; amount = amt; 
    req_time = block.number
  }
  
  function finalize() {
    require(state==1); state = 0; // REQ -> IDLE
    require(block.number>=req_time+wait_time);
    require(msg.sender==owner);	
    receiver.transfer(amount)
  }
  
  function cancel() {
    require(msg.sender==recovery);
    require(state==1); state = 0 // REQ -> IDLE
  }
}
\end{lstlisting}
	\end{figure}
}

This contract allows the owner to deposit and withdraw tokens, with a mechanism to cancel pending withdrawal requests through a recovery key. 
Its behaviour is modelled as an automaton, \ie certain actions are enabled only in certain states and at certain times.
Once the vault contract (see \Cref{lst:vault}) have been deployed, it supports the following actions:
\begin{inlinelist}
\item \code{withdraw}, which allows the owner to issue a withdraw request to the vault, specifying the receiver and the desired amount;
\item \code{finalize}, which allows the owner to finalize the withdraw after the wait time has passed since the request;
\item \code{cancel}, which allows the owner of the recovery key to cancel the withdraw request during the wait time.
\end{inlinelist}
The action \code{withdraw} implements a state transition from IDLE to REQ, while \code{finalize} and \code{cancel} from REQ to IDLE. 

}{}

\section{\ourlogic: a temporal logic for contracts}
\label{sec:logic}

We now introduce our property specification language.
Its foundation lies in Hennessy-Milner logic (HML)~\cite{HennessyMilner85jacm}, 
a modal logic interpreted over labelled transition systems (LTSs), with a modal construct $\txnext{a}\,\phi$ to express the ``possibility'' of satisfying a state formula $\phi$ after performing the action $a$.
More precisely, a state satisfies $\txnext{a}\,\phi$ if it admits an $a$-labelled transition to a state satisfying $\phi$. 

Interpreting HML over the LTS of contracts in~\Cref{sec:contracts} would allow us to specify properties such as:
``after firing transaction $\txT$, the balance of $\pmvA$ is $n$''.
However, specifying precise conditions on the reached state, 
such as ``the balance of $\pmvA$ is $n$'', is not particularly meaningful in our context. 
Rather, one is typically interested in relational properties between the pre- and post-state, such as ``the balance of $\pmvA$ is increased by $n$''.
To capture this, we interpret formulae over \emph{sequences} of blockchain states, and we introduce a construct $\oldL{e}$ that allows evaluating an expression $e$ in the state before the last transition.

Note however that the object referred to by the modal operator of HML is a \emph{constant}. 
Consequently, applying HML to specify contract properties would require to explicitly provide the \emph{exact} transaction $\txT$ that leads to the new state.
This would rule out the specification of more general and interesting properties such as:
``Can $\pmvA$ fire some transaction that increases her balance?'', or
``Does there exist some address $\pmvB \neq \pmvA$ that can fire a transaction decreasing $\pmvA$'s balance?''.

To overcome this limitation, we consider a first-order version of HML and  extend the syntax of the modal operator to allow constraints over the individual fields of a transaction.
The general form of our modal operator is the following:
\[
\txnext{\txL{\pmvColor{e_s}}{\contract{C}}{e_f}{e_1,\ldots,e_k}{e_v}} \, \phi
\]
where $\pmvColor{e_s}$ is an expression specifying the transaction sender,
$\contract{C}$ is the target contract (a constant address), $e_f$ is the called procedure (either constant or variable), $e_1,\ldots,e_k$ is a sequence of expressions representing the procedure parameters, and $e_v$ is an expression specifying the value transferred along with the transaction.
Since we are adopting a first-order version of HML, variables occurring in the modal operator can be universally or existentially quantified. 
This makes it possible to express complex properties like those informally discussed earlier.  



\iftoggle{arxiv}{%
\subsection{Syntax}
}
{%
\myparagraph{Syntax}
}
\label{sec:logic:syntax}
We define properties of blockchain states as formulae $\phi, \phi', \ldots$, with the syntax in~\Cref{fig:logic:syntax}.
The base case is given by (boolean) expressions on blockchain states,
which extend those of the contract language with the constructs $\oldL{e}$ and $\lastrevL$.
Intuitively, $\oldL{e}$ evaluates the expression $e$ in the previous computation state, while $\lastrevL$ is true whenever the transaction leading to the current state \emph{reverted} (\ie, it is not valid).

Logical negation and conjunction are standard; the other connectives can be obtained as syntactic sugar as usual.
The formula $\forall x:\typeT.\ \phi$ universally quantifies over the variable $x$ (of type $\typeT$).
As usual, $\exists x:\typeT.\ \phi$ is syntactic sugar for 
$\neg \forall x:\typeT.\, \neg \phi$.
The formula 
$\txnext{\txL{\pmvColor{e_s}}{\contract{C}}{e_f}{e_1,\ldots,e_k}{e_v}} \, \phi$
generalises the modal operator $\txnext{a}$ of HML in order to allow a fine-grained treatment of transaction parameters. 

\iftoggle{arxiv}{}
{%
\ourtool includes a type checker that prevents typing errors.
For example, the formula 
$\phi_1 = \walE{\contract{C}} > \oldL{\walE{\contract{C}}}$
is rejected by the type checker, because the \code{old} cannot refer to a previous state.
Instead, 
$\txnext{\txL{\pmvA}{\contract{C}}{\procF}{}{0}} \, \phi_1$
is well-typed whenever the contract $\contract{C}$ has a method $\txcode{f}$ with zero parameters.
Because of space constraints, we describe the typing rules of \ourlogic in~\Cref{app:logic}.
}

\begin{figure}[t]
\centering
\footnotesize
\resizebox{\linewidth}{!}{
\fbox{
\begin{minipage}{1.06\linewidth}
\vspace{-10pt}
\hspace{-15pt}
\begin{minipage}{0.48\linewidth}
\[
\begin{array}{rcll}
e 
& \bnfdef
& 
& \textbf{Expression}
\\
& \bnfmid 
& n
& \text{integer constant}
\\
& \bnfmid 
& \addrA
& \text{address constant}
\\
& \bnfmid
& \procF 
& \text{procedure constant}
\\
& \bnfmid 
& x
& \text{variable}
\\
& \bnfmid 
& e.x 
& \text{field}
\\
& \bnfmid 
& e.x[e'] 
& \text{map entry}
\\
& \bnfmid 
& \senderE 
& \text{transaction sender}
\\
& \bnfmid 
& \valueE 
& \text{transaction value}
\\
& \bnfmid
& e_1 \circ e_2
& \text{operation}
\\
& \bnfmid 
& \oldL{e}
& \text{previous state}
\\
& \bnfmid 
& \lastrevL\;\;\;
& \text{transaction revert}
\end{array}
\]
\end{minipage}
\begin{minipage}{0.52\linewidth}
\[
\begin{array}{rcll}
\typeT
& \bnfdef
& 
& \textbf{Type}
\\
& \bnfmid
& \boolT
& \text{bool}
\\
& \bnfmid
& \intT
& \text{integer}
\\
& \bnfmid
& \addressT
& \text{address}
\\
& \bnfmid
& \methodT
& \text{method}
\\
& \bnfmid
& \calldataargsT
& \text{any params}
\\[5pt]
\phi 
& \bnfdef
& 
& \textbf{Formula}
\\
& \bnfmid
& e
& \text{expression}
\\
& \bnfmid
& \neg\phi
& \text{negation}
\\
& \bnfmid
& \phi_1 \land \phi_2
& \text{conjunction}
\\
& \bnfmid
& \forall x \hastype \typeT. \; \phi 
& \text{quantification}
\\
& \bnfmid
& \txnext{\txL{\pmvColor{e_s}}{\contract{C}}{e_f}{e_1,\ldots,e_k}{e_{v}}} \, \phi \;\;\;
& \text{transaction}
\end{array}
\]
\end{minipage}
\end{minipage}
}
}
\caption{Syntax of contract specifications.}
\label{fig:logic:syntax}
\end{figure}

\iftoggle{arxiv}{%
\iftoggle{arxiv}{%
\subsection{Typing of \ourlogic logic}
}
{%
\section{Typing}
\label{app:logic}
}

While in~\Cref{sec:contracts} we followed a minimalist approach, presenting an untyped version of the contract language, here we opt for a typed presentation of the specification language, since the use of types therein is not as straightforward.
However, for succinctness we only elaborate on the less standard aspects of our type system.
\ourtool includes a type checker that ensures the well-typedness of specifications and prevents typing errors.

The types $\typeT$ (see~\Cref{fig:contracts:syntax}) include the base types $\boolT$, $\intT$ and $\addressT$ already (implicitly) used in the contract language, and the additional types $\methodT$ and $\calldataargsT$.
The type $\methodT$ represents the set of all procedure names, 
while $\calldataargsT$ represents the set of all possible \emph{sequences} of procedure parameters.
These types are typically used in synergy with quantification and the modal operator to specify \emph{transition invariants}, \ie properties that must hold for every possible procedure call.
For example, we can express the property
``for all state transitions, the balance of $\pmvA$ is preserved''
as follows:
\begin{align*}
& \forall a \hastype \addressT.\;
\forall f \hastype \methodT.\;
\forall x \hastype \calldataargsT.\;
\forall n \hastype \intT. \\
& \qquad
\txnext{{\txL{a}{\contract{C}}{f}{x}{n}}} \;\;
\walE{\pmvA} = \oldL{\walE{\pmvA}}
\end{align*}

We rule out ill-typed specifications through a type system:
its rules are mostly standard, with the only exception of the 
rules to ensure that $\oldL{e}$ expressions do not go out of scope.
To this purpose, type judgments for expressions have the form 
$\Gamma,n \vdash e : \typeT$, where $\Gamma$ is a mapping from variables to types and $n$ is a nonnegative integer representing the level of nesting of $e$ within modal operators.
The key typing rule for expressions is:
\[
\irule
    {\Gamma, n-1 \vdash e : \typeT}
    {\Gamma, n \vdash \oldL{e} : \typeT}
\;\; n>0
\]
The rest of the typing rules for expressions are standard. 

Typing judgments for formulae have the form $\Gamma,n \vdash \phi$.
We say that $\phi$ is well-typed when $\emptyset,0 \vdash \phi$, where $\emptyset$ is the empty typing environment. 
The key typing rules are those for quantified formulae and 
for the modal operator:
\[
\begin{array}{c}
\irule
    {\Gamma\setenum{\bind{x}{\typeT}},n \vdash \phi}
    {\Gamma,n \vdash \forall x:\typeT. \phi}
\\[15pt]
\irule
    {\begin{array}{c}
    \Gamma,n \vdash e_s \hastype \addressT \quad 
    \Gamma,n \vdash e_v \hastype \intT
    \quad
    \Gamma,n+1 \vdash \phi
    \\[2pt]
    \procF(x_1:\typeT[1],\ldots,x_k:\typeT[k]) \in \contract{C}
    \quad
    \forall i \in 1..k : \;\; 
    \Gamma,n \vdash e_i \hastype \typeT[i]
    \end{array}
    }
    {\Gamma,n \vdash \txnext{\txL{\pmvColor{e_s}}{\contract{C}}{\procF}{e_1,\ldots,e_k}{e_{v}}} \, \phi}
\\[15pt]
\irule
    {\begin{array}{c}
    \Gamma,n \vdash e_s \hastype \addressT \quad 
    \Gamma,n \vdash e_v \hastype \intT
    \quad
    \Gamma,n+1 \vdash \phi
    \\[2pt]
    \Gamma,n \vdash x \hastype \methodT \quad
    \Gamma,n \vdash y \hastype \calldataargsT
    \end{array}
    }
    {\Gamma,n \vdash \txnext{\txL{\pmvColor{e_s}}{\contract{C}}{x}{y}{e_{v}}} \, \phi}
\end{array}
\]

The rule for typing universal quantifications is straightforward: 
$\forall x:\typeT. \phi$ is typeable in $\Gamma$ if 
$\phi$ is typeable in a typing environment that extends $\Gamma$ with the 
binding from $x$ to $\typeT$.
We split the typing of the modal operator in two rules for convenience.
In the first rule, $\procF$ is a constant function name.
In that case, the premises require all the function parameters to have the type declared in the contract (here we assume a typed version of the contract language).
Note that $\phi$ must be typeable in a judgment with $n+1$, since we are adding a nesting level.
The third rule deals with the case where the function name is a variable $x$: in that case, the only possible parameter can be another variable $y$, which must have type $\calldataargsT$.






For example, let $\contract{C}$ be a contract with a procedure $\procF$ with zero parameters and a procedure $\procG$ with one $\intT$ parameter, 
and consider the formulae: 
\begin{align*}
\phi_1 
& = \walE{\contract{C}} > \oldL{\walE{\contract{C}}}
\\
\phi_2 
& = \txnext{\txL{\pmvA}{\contract{C}}{\procF}{}{0}} \, \phi_1
\\
\phi_3 
& = \forall x \hastype \methodT.\ \forall y \hastype \calldataargsT.\ \txnext{\txL{\pmvA}{\contract{C}}{x}{y}{0}} \, \phi_1
\end{align*}
The formula $\phi_1$ is not well-typed, since inferring a typing judgment  
$\emptyset,0 \vdash \oldL{\walE{\contract{C}}} \hastype \intT$ would require to satisfy the premise
$\emptyset,-1 \vdash \walE{\contract{C}} \hastype \intT$, which is forbidden by the rule premises. 
Instead, both $\phi_2$ and $\phi_3$ are well-typed, since the rule for typing the modal operator increases the $n$ in the typing judgment, 
and $\emptyset,1 \vdash \phi_1$.
}
{}

\iftoggle{arxiv}{%
\subsection{Semantics}
}
{%
\myparagraph{Semantics}
}
\label{sec:logic:semantics}
To interpret \ourlogic, we first extend the contract semantics of~\Cref{sec:contracts} in order to have a left-total transition relation, admitting state transitions for all possible labels.
The intuition is to flag blockchain states $\st$ to indicate whether the last fired transaction was enabled or not. 
Formally, the states $\sysS$ of this extended system are pairs $(\st,b)$ where $\st$ is a blockchain state and $b$ is a bit representing whether the last transaction was reverted ($b=1$) or not ($b=0$).
We define the transition relation $\xmapsto{\;\;}$ between flagged blockchain states as follows:
\[
\irule
    {\st \xrightarrow{\;\txT\;} \sti}
    {(\st,b) \xmapsto{\;\txT\;} (\sti,0)}
\qquad
\irule
    {\st \;\not\!\!\xrightarrow{\;\txT\;}}
    {(\st,b) \xmapsto{\;\txT\;} (\st,1)}
\]
The relation $\xmapsto{\;\;}$ is left-total: for all $\sysS$ and $\txT$, there exists $\sysSi$ such that $\sysS \xmapsto{\txT} \sysSi$. 

We interpret expressions and formulas over non-empty finite sequences $\sysSS $ of flagged states.
Given a (possibly empty) sequence $\sysSS$ and a state $\sysS$,  
we denote by $\sysS::\sysSS$ the sequence with head $\sysS$ and tail $\sysSS$, with $\hd{\sysSS}$ the head of $\sysSS$, and with $\tl{\sysSS}$ its tail (which is possibly empty).

The interpretation of expressions in contract specifications follows closely that in~\Cref{fig:contracts:sem-exp}.
We only elaborate on the semantics of the new expressions, 
\ie $\oldL{e}$ and $\lastrevL$.
The evaluation of $\oldL{e}$ in a sequence $\sysSS$ is defined as
\(
\semexp[\sysSS]{\oldL{e}} = \semexp[\tl{\sysSS}]{e}
\),
\ie the evaluation of $e$ in the previous step of the computation $\tl{\sysSS}$.
The evaluation of $\lastrevL$ just retrieves the revert bit from the heading flagged state, \ie
\(
\semexp[\sysSS]{\lastrevL} = b
\)
whenever $\hd{\sysSS} = (\st,b)$.

The interpretation of formulae is defined in~\Cref{fig:logic:semantics}.
In the first rule, note that the type system ensures that only boolean expressions can be interpreted.
The rules for logical connectives are standard.
For universal quantification, we have two rules.
The first rule deals with variables of any type except $\calldataargsT$: we substitute the variable $x$ with any possible value of the same type, and require that the resulting formula is true.
Note that these substitutions preserve well-typedness.
The second rule deals with the case where $x$ has type $\calldataargsT$:
we substitute $x$ in $\phi$ with any possible sequence of values $\vec{v}$, provided that the resulting formula is well-typed.
The rule for the modal operator requires the transaction $\txT$ to be fully instantiated (this is guaranteed by the type system).
It first computes the new state $\sysS$ reached upon firing $\txT$ in the current state $\hd{\sysSS}$, and then evaluates $\phi$ in $\sysS::\sysSS$.
Since $\xmapsto{\;\;}$ is left-total, this rule is always applicable.

\begin{figure}[t]
	\input{fig_logSem}
\end{figure}

Our verification goal is to decide, given a formula $\phi$ and an initial state $\sysS$, whether $\sysSi \models \phi$ for all $\sysSi$ reachable from $\sysS$. \enriconote{questo secondo me va chiarito molto prima, già nel cappello}

\iftoggle{arxiv}{%
\subsection{Example: a bet contract}
}
{%
\myparagraph{Example: a bet contract}
}
\label{sec:logic:example}
Consider a $\contract{Bet}$ contract with the following methods:
\begin{inlinelist}
\item the constructor sets an oracle and the initial bet, which is the amount of tokens transferred upon deployment;
\item \code{join} allows anyone to join the bet, by depositing into the contract an amount of tokens equal to the initial bet;
\item \code{win} allows the previously joined player to withdraw the whole pot, provided that \code{rate} is greater than 100;
\item \code{set} allows the oracle to set the rate.
\end{inlinelist}
 
We show how to specify relevant properties of the bet contract in \ourlogic.
Consider the property:
``if the \texttt{rate} is greater than 100 and the player has joined, then some user can fire some transaction to withdraw the entire pot''.
We call this property \textbf{winnability} and formalise it in \ourlogic as follows:
\begin{align*}
\phi_1 = \;
& (
\contract{Bet}.\texttt{rate}>100 \,\land\, \contract{Bet}.\texttt{player}\neq \nullE
)
\rightarrow
\big( 
\exists a \hastype \addressT .\
\exists f \hastype \methodT .\
\exists x \hastype \calldataargsT .\
\\
& \txnext{{\txL{a}{\contract{Bet}}{f}{x}{0}}} \;\;
\walE{a} = \oldL{\walE{a} + \walE{\contract{Bet}}}
\big)
\end{align*}


We expect $\phi_1$ to be true in all reachable states: the procedure to be called is $\txcode{win}$, and the only possible sender is $\contract{Bet}.\texttt{player}$. 
Note instead that if we replace the $\exists a$ with a $\forall a$, \ie if we ask whether \emph{any} user can withdraw the pot, then the property becomes false, despite the fact that anyone can call $\txcode{win}$.

A general desirable property is \textbf{liquidity}, \ie it is always possible to withdraw all tokens from the contract.
For the $\contract{Bet}$ contract, it is always possible to achieve this goal in at most two steps.
We formalise this property in \ourlogic as follows:
\begin{align*}
\phi_2 = \;
&  
\contract{Bet}.\texttt{player}\neq \nullE
\rightarrow  
\big(
\exists a_1,a_2 \hastype \addressT .\
\exists f_1,f_2 \hastype \methodT .\
\exists x_1,x_2 \hastype \calldataargsT .\
\\
& 
\txnext{{\txL{a_1}{\contract{Bet}}{f_1}{x_1}{0}}} \;
\txnext{{\txL{a_2}{\contract{Bet}}{f_2}{x_2}{0}}} \;
\walE{\contract{Bet}} = 0
\big)
\end{align*}


We now ask whether the contract is subject to   
attacks in which an adversary front-runs another user's transaction in order to prevent the user from winning.
In the $\contract{Bet}$ contract, we have such an attack when the player is willing to fire a $\txcode{win}$ transaction, and the adversary front-runs it with a transaction that decreases the \texttt{rate}, making the player's transaction revert.
We formalise this \textbf{frontrunning} attack in \ourlogic as follows:
\begin{align*}
\phi_3 = \;
&  
\forall a \hastype \addressT .\
\exists b \hastype \addressT.\
\exists f \hastype \methodT .\
\exists x \hastype \calldataargsT .\
\\
& 
\txnext{{\txL{b}{\contract{Bet}}{f}{x}{0}}} \;
\txnext{{\txL{a}{\contract{Bet}}{\txcode{win}}{}{0}}} \;
\lastrevL
\end{align*}

We expect $\phi_3$ to be true: indeed, any $\txcode{win}$ transaction is reverted when 
$\contract{Bet}.\texttt{oracle}$ anticipates it with a transaction $\txcode{set}\texttt{(x)}$ with argument $\texttt{x} \leq 100$.
If we modify $\phi_3$ by requiring that $b \neq \contract{Bet}.\texttt{oracle}$, we expect instead the property to be false, since only the oracle has the rights to call $\txcode{set}$.


\section{Encoding in first-order logic}
\label{sec:encoding}

We now formalize our encoding from contracts and \ourlogic into first-order logic, and refer to~\Cref{app:encodinf-fol:bet}  
for a worked example.


We assume a set of state variables ($\StateVars$) and a set of transition variables ($\TransVars$), with types ranging over
$\boolT$, $\intT$, $\addressT$, $\methodT$, or ${\addressT \Rightarrow \tau}$ with $\tau \in \setenum{\boolT, \intT, \addressT}$.
Let $\Addresses = \setcomp{v}{\text{type}{(v)} = \type{address}}$ be a finite set,
let ${\Contracts = \setenum{C_1 \dots, C_n}} \subseteq \Addresses$ the set of contract addresses,
and
let $\Methods = \setenum{f_1, \dots, f_p}$ the set of 
constants of type $\methodT$ representing procedures.

State variables include the following:
\begin{itemize*}[label=]

\item $\balanceSMT \hastype \addressT \to \intT$ representing address balances,

\item $\lastRevert \hastype \boolT$ representing whether the last transaction reverted,

\item $\constructed_{C} \hastype \boolT$ for every $C \in \Contracts$, representing whether $C$ has already been constructed or not, 

\item and a variable for each field of each contract, with its given type. We denote the set of such variables as \FieldsVars.

\end{itemize*}

Transition variables include:
\begin{itemize*}[label=]
\item $\xavar \hastype \addressT$ for the transaction sender,

\item $\xnvar \hastype \intT$ for the transaction value,

\item $\txcontract \hastype \addressT$ for the called contract,

\item $\txfunction \hastype \methodT$ for the name of the called procedure,
\item and a variable for each parameter of each procedure, with its respective type. Given a constant $f\in\Methods$ representing a procedure with parameters $x_1, \dots, x_k$, we denote the set of variables corresponding to these parameters as $\ParamVars(f)=\{f\_x_1, \dots, f\_{x_k}\}$. 
We also define $\ParamVars \defas \bigcup_{f\in \Methods} \ParamVars(f)$.
\end{itemize*}
Given these definitions, the encoding of expressions is straight-forward. Occurrences of \thisE are replaced by the address constant that respresents the contract in which \thisE appears.  

\iftoggle{arxiv}{%
\subsection{Encoding the contract}
}
{%
\myparagraph{Encoding the contract}
}
\label{sec:encoding:contract}
We now show how to encode the state transition system of \Cref{sec:contracts} into first-order logic formulas.
As common in SMT-based model checking~\cite{SATbasedModelChecking}, we define a formula $\Init$ that describes the set of initial states, 
and a formula $\Trans$ that describes the relation between current state variables, transition variables, and next state variables.

$\Init$ is the conjunction of the following constraints:
\begin{itemize*}[label=]

\item $\bigwedge_{a \in \Addresses} \balanceSMT[a] \geq 0$ (balances are nonnegative),


\item $\bigwedge_{c \in \Contracts} \constructed_c = \falsesmt$ (contracts have not been constructed yet),

\item and $\bigwedge_{v \in  \FieldsVars} v = \defvalue(\typeV(v))$ (all contract fields are set to a default value determined by their type).

\end{itemize*}

Since the imperative nature of the contract language
allows variables to be updated multiple times within a single call,
to capture intermediate states in the definition of $\Trans$
we introduce sets of temporary state variables 
$\StateVars^j$ for $j \in \range{1}{m+1}$
 --- where $m$ is the maximum  depth  of statements --- 
each $\StateVars^i$ being a copy of $\StateVars$. 
For all $v\in \StateVars$, we denote with $v^i$ the corresponding temporary variable in $\StateVars^i$. 
Finally, we introduce a copy \NextVars of $\StateVars$ to represent ``next'' state variables, and, for all $v\in \StateVars$, we denote with $v^{nx}$ the corresponding variable in \NextVars.
The formula \Trans describes the relation between \StateVars, \TransVars, and \NextVars.

%

We now define the encoding of 
a procedure 
\(
\procF(x_1,\ldots,x_n) \braceleft s_1; \dots; s_{m} \braceright
\)
in a contract $C$.
W.l.o.g., we assume that none of the $s_i$ 
contains sequential compositions.
Initially, the nesting level is $i=0$, and we define ${\StateVars^0 \defas \StateVars}$.
%
We define an operator $\encS$  that maps 
statements to first-order formulas, 
such that, for each $s_i$, $\encS(s_i)$ uniquely determines the values of all the variables in $\StateVars^{i+1}$
as a function of the values of the variables in  $\StateVars^{i}$ and in $\TransVars$. 
For brevity, in the description below we omit the constraints of the form $v^{i+1} = v^i$, \ie, when the value of $v$ does not change passing from nesting level $i$ to $i+1$, we do not explicitly write the equality.



First, we encode the transfer of $\xnvar$ tokens from $\xavar$ to $C$ made along with the call (which can be seen as an implicit default statement $s_0$):
\begin{align*}
\encS(s_0) \defas  
\ITE(&send\_enabled^0(\xavar, C, \xnvar),
\\ 
& \lastRevert^{1} = \falsesmt 
\land send^0(\xavar, C, \xnvar) ,		
\\ 
& 
\lastRevert^{1} = \truesmt
)
\end{align*}
where ITE is the standard if-then-else construct, and:
\iftoggle{arxiv}{
	\begin{align*}
		& send\_enabled^i(e_0, e_1, e_2) 
		\defas \balanceSMT[e_0]^i  \geq e_2^{i}  \land e_0^i \neq e_1^i
	\end{align*}
	and
	\begin{align*}
		send^i(e_0, e_1, e_2) 
		\defas 
		\bigwedge_{a\in \Addresses}   \balanceSMT[a]^{i+1} =  
		& 
		\ITE\big(
		\, a = e_0^{i},  \\
		& \qquad \
		\balanceSMT[a]^{i} - e_2^{i},    
		\, \\
		 & \qquad \
		\ITE( 
			a = e_1^i,  \\
			& \qquad \quad \quad \ \balanceSMT[a]^{i} + e_2^{i},\\
			& \qquad \quad \quad \ 
			\balanceSMT[a]^{i}
			)
		\big)
	\end{align*}
}
{
	\begin{align*}
	& send\_enabled^i(e_0, e_1, e_2) 
	\defas \balanceSMT[e_0]^i  \geq e_2^{i}  \land e_0^i \neq e_1^i
	\\
	& send^i(e_0, e_1, e_2) 
	\defas 
	\bigwedge_{a\in \Addresses}   \balanceSMT[a]^{i+1} = 
	\\  
	& 
	\qquad 
	\ITE\big(
	\, a = e_0^{i}, 
	\,
	\balanceSMT[a]^{i} - e_2^{i},    
	\, 
	\ITE( 
	a = e_1^i, 
	\balanceSMT[a]^{i} + e_2^{i},
	\balanceSMT[a]^{i}
	)
	\big)
	\end{align*}
}

\begin{table*}[t]
\centering
\caption{Encoding of contracts into first-order logic.}
\label{tab:statement-to-fol}
\begin{tabular}{|c|l|}
\hline
\textbf{Statement} $s_i$ & \multicolumn{1}{c|}{\textbf{FOL encoding} $\encS(s_i)$}
\\
\hline
$\skipE$ 
&
$\truesmt$
\\[2pt]
$\reqE{e}$
&
$\lastRevert^{i+1} = \ITE(e^{i}, \falsesmt, \truesmt)$
\\[2pt]
$x = e$
&
$x^{i+1} = e^{i}$
\\[2pt]
$x[e_1] = e_2$
&
$\bigwedge_{a\in \Addresses} x(a)^{i+1} = \ITE(a = e_1^{i}, e_2^{i}, x\_a^{i})$
\\[2pt]
$\ifE{e}{s_{i}^{1}}{s_{i}^{0}}$
&
$\ITE(e^{i}, \encS(s_{i}^{1}), \encS(s_{i}^{0}))$
\\
$\txout{e_1}{e_2}{}$
&
$\ITE(send\_enabled^i(C, e_1, e_2),
    \begin{array}{l}
    \lastRevert^{i+1} = \falsesmt \land send^i(C, e_1, e_2),\\ 
	\lastRevert^{i+1} = \truesmt)
    \end{array}$
\\
\hline
\end{tabular}
\end{table*}

\noindent
Then, for each statement $s_i$, we define the encoding $\encS(s_i)$ in~\Cref{tab:statement-to-fol}. 


A procedure reverts if any of its statements $s_i$ reverts, \ie if $\lastRevert^i = \truesmt$ for some $i \in \range{1}{m+1}$.
If it does not revert, then the next state of each variable $v$ is given by $v^{m+1}$.
Overall, for a procedure $p$, the relation between current state variables and next state variables is given by:
\begin{align*}  
\Trans_p \defas
\ITE\,\Big(
& \bigvee_{i \in \range{1}{m+1}} \!\!\!\lastRevert^i,
\bigwedge_{v\in \StateVars} \!\!\! v^{nx} = v^{0},
\bigwedge_{v\in \StateVars} \!\!\! v^{nx} = v^{m+1} \land \bigwedge_{i \in \range{0}{m}} \!\!\! \encS(s_i) 
\Big)
\end{align*}


Note that $\Trans_p$ depends on all the $\StateVars^j$,
for $j \in \range{0}{m+1}$. However, by the definition of $\encS^i$ given before, we have that 
for all $i \in \range{1}{m+1}$ and $v^i \in \StateVars^i$,
there is an equation of the form $v^i = \psi_v^i(\StateVars^{i-1}, \TransVars)$,
for a suitable $\psi_v^i$ that only depends on variables in 
$\StateVars^{i-1}$ and in 
$\TransVars$.
Hence, we can rewrite the equations in $\Trans_p$ in the form
$v^{nx} = \chi_v(\StateVars, \TransVars)$, 
where each $\chi_v(\StateVars, \TransVars)$ is obtained by
exhaustively substituting, for every variable $v\in\StateVars$ and  for every $i \in \range{1}{m+1}$, all the occurrences of $v^i$ with  $\psi_v^i(\StateVars^{i-1}, \TransVars)$.
After such substitutions, we obtain that $\Trans_p$ only depends on $\StateVars$, $\TransVars$, and $\NextVars$.

%

In order to encode a contract
$c \equiv \contract{C} \; \braceleft \; ({p_1}, \dots, {p_k}) \; \braceright$,
we simply define
\iftoggle{arxiv}{
	$${\Trans_c \defas \switch\ \txfunction\ \{ \mycase\ p_1 : \Trans_{p_1}\ \mid \cdots \mid \mycase\ p_k : \Trans_{p_k}\}}$$ 
}
{
	${\Trans_c \defas \switch\ \txfunction\ \{ \mycase\ p_1 : \Trans_{p_1}\ \mid \cdots \mid \mycase\ p_k : \Trans_{p_k}\}}$, 
}
where the switch-case operator is just syntactic sugar for a sequence of nested $\ITE$s.
The encoding of a whole system is:

\iftoggle{arxiv}{
	$${\Trans \defas \switch\ \txcontract\ \{\, \mycase\ c_1 : \Trans_{c_1} \mid \cdots \mid \mycase\ c_k : \Trans_{c_k}\}}$$
}
{
	${\Trans \defas \switch\ \txcontract\ \{\, \mycase\ c_1 : \Trans_{c_1} \mid \cdots \mid \mycase\ c_k : \Trans_{c_k}\}}$.
}


\iftoggle{arxiv}{%
\subsection{Encoding the property}
}
{%
\myparagraph{Encoding the property}
}
\label{sec:encoding:property}
%
%
We now show how to encode \ourlogic into first-order logic. 
Such encoding is rather straightforward except for the modal operator 
$\txnext{\cdot}$,
for the operator $\oldL{\cdot}$,
and for the  type \calldataargsT.

First, note that the nesting of modal operators within a formula $\phi$ gives rise to a tree \Tree, defined as follows.
We associate to $\phi$ the root of the tree.
Then, recursively, for every occurrence of the modal operator $\txnext{a}\psi$, we associate to the subformula $\psi$  a child of the current node, and to $a$ the connecting edge.

To each node \node of the tree, we associate  a copy of the set of state variables (denoted by $\StateVars^{\node}$), and to each edge \edge we associate a copy of the set of transition variables ($\TransVars^\edge$).
For $v\in \StateVars$, 
we denote with $v^\node$ the corresponding fresh variable in $\StateVars^\node$, 
and, similarly, 
for $v\in \TransVars$, 
we denote with $v^\edge$ the corresponding fresh variable in $\TransVars^\edge$. 
%

We define an operator $\encP^{\node,\edge}$ that, given a node $\node$ and an edge $\edge$, maps expressions and formulae of the property specification language to first-order logic terms and formulae.

For every formula $\phi$ not containing the modal operator, the \oldL{} operator, or variables of type \calldataargsT, we define
$
\encP^{\node,\edge}(\phi) \defas  \phi\setenum{\bind{\StateVars}{\StateVars^\node}}\setenum{\bind{\TransVars}{\TransVars^\edge}}
$.


Now we consider the remaining cases.
\sloppy{Let $\phi$ be a formula containing the subformula  
${\txnext{\txL{\pmvColor{e_s}}{\contract{C}}{e_f}{e_1,\ldots,e_k}{e_{v}}}}\psi$,
and let $\node$  be the node associated to $\phi$,
 \edge be the edge associated to $\txnext{\txL{\pmvColor{e_s}}{\contract{C}}{e_f}{e_1,\ldots,e_k}{e_{v}}}\psi$,
and \nodei be the child of \node through \edge (\ie the node associated to $\psi$).}
Then, we define:
%
%
%

\iftoggle{arxiv}{
	\begin{align*}
	&\encP^{\node,\edge}{(\txnext{\txL{\pmvColor{e_s}}{\contract{C}}{e_f}{e_1,\ldots,e_k}{e_{v}}} \ \psi)} \defas
	\; \\
	& \qquad \Exists \StateVars^\nodei \ .  \\
	& \qquad  \Exists \TransVars^\edge \ . 
	\\
	& \qquad \Trans
	\setenum{\bind{\StateVars}{\StateVars^\node}}
	\setenum{\bind{\TransVars}{\TransVars^\edge}}
	\setenum{\bind{\NextVars}{\StateVars^\nodei}}
	\\ 
	& \qquad \land  \xavar^\edge = \encP^{\node,\edge}({\pmvColor{e_s}})  
	\\ & \qquad 	   \land    \txcontract^\edge = \contract{C}
	 \\ & \qquad \land \txfunction^\edge = \encP^{\node,\edge}({e_f}) 
	\\  
	& \qquad \land  \EqParams
	\\ 
	& \qquad \land   \xnvar^\edge = \encP^{\node,\edge}({e_v}) 
	\\  
	& \qquad \land \encP^{\nodei,\edge}{(\psi)}
	\end{align*}
}
{
	\begin{align*}
	&\encP^{\node,\edge}{(\txnext{\txL{\pmvColor{e_s}}{\contract{C}}{e_f}{e_1,\ldots,e_k}{e_{v}}} \ \psi)} \defas
	\;
	\Exists \StateVars^\nodei \ .   \Exists \TransVars^\edge \ . 
	\\
	& \qquad \Trans
	\setenum{\bind{\StateVars}{\StateVars^\node}}
	\setenum{\bind{\TransVars}{\TransVars^\edge}}
	\setenum{\bind{\NextVars}{\StateVars^\nodei}}
	\\ 
	& \qquad \land \;\; \xavar^\edge = \encP^{\node,\edge}({\pmvColor{e_s}}) 
	\quad  \land \quad   \txcontract^\edge = \contract{C}
	\quad \land \quad  \txfunction^\edge = \encP^{\node,\edge}({e_f}) 
	\\  
	& \qquad \land \;\;  \EqParams
	\quad \land  \quad \xnvar^\edge = \encP^{\node,\edge}({e_v}) 
	\quad \land \quad \encP^{\nodei,\edge}{(\psi)}
	\end{align*}
}
where, in the case in which $e_f$ is a procedure constant with $k$ parameters (\ie $e_f \equiv f$, for  $f\in \Methods$), 
$
\EqParams \defas  \bigwedge_{v_i\in\ParamVars(f)} v_i^\edge = \encP^{\node,\edge}({e_i}) 
$.
Otherwise, the only other case admitted by the \ourlogic type checker is that $e_f$ is a variable with a single argument of type \calldataargsT, 
\ie $e_f(e_1,\ldots,e_k) \equiv x(arg)$, with $x\ :\ \methodT$ and $arg\ :\ \calldataargsT$. 
In order to define $\EqParams$ in this second case, we explain how variables of the type $\calldataargsT$ are dealt with in the encoding in first-order logic.

The type \calldataargsT, from the perspective of the encoding in first-order logic, can be seen as compact way to allow reasoning about the set of the procedure parameters \ParamVars inside the arguments of the modal operators.
We associate to each variable $arg\ :\ \calldataargsT$  a copy $\ParamVars^{arg}$ of $\ParamVars$ (and, for all $v\in \ParamVars$, we denote with $v^{arg}$ the corresponding variable in $\ParamVars^{arg}$). 
Then, every quantification on $arg$ corresponds to a quantification on all the variables in $\ParamVars^{arg}$. That is, for  the $\forall$ case (the $\exists$ case is analogous):
\[
 \encP^{\node,\edge}(\forall arg.\ \phi ) 
 \defas
	\Big(
	\Forall_{v\in \ParamVars} v^{arg}
	\Big)
	.\
	 \encP^{\node,\edge}(\phi)
\]  
We then define
$\EqParams \defas \bigwedge_{v\in\ParamVars} v^\edge = v^{arg}$.
Finally, if we are in a node $\node$, with $\node_p$ being the parent of $V$, we define
\[
\encP^{\node,\edge}
	(\oldL{\phi}) 
		\defas 
	\encP^{\node_p,\edge}(\phi)
\]
Note that the existence of the parent node $\node_p$, for a node \node that contains an \oldL{} expression is guaranteed by the type checker.


\iftoggle{arxiv}{\iftoggle{arxiv}{%
\subsection{Encoding of the Bet contract}
}
{%
\section{First-order logic encoding of the Bet contract}
}
\label{app:encodinf-fol:bet}

In this~\namecref{app:encodinf-fol:bet} we exemplify our encoding into first-order logic by applying it to our Bet contract and its winnability property.

\iftoggle{arxiv}{%
	\subsubsection{Encoding the contract.}
}
{
	\subsection{Encoding the contract}
}

Recall the contract $\contract{Bet}$ in~\Cref{lst:bet}.
We have:
\begin{itemize}

\item $\Contracts=\setenum{Bet}$

\item $\Methods = \setenum{constructor, pay, join, set}$

\item $\StateVars = \{\balanceSMT, \lastRevert,\\ \constructed_{Bet}, oracle, rate, player\}$, where:
\begin{itemize}
    \item $oracle, player \hastype \addressT$
    \item $rate \hastype \intT$
\end{itemize}  

\item $\TransVars = \{\xavar, \xnvar, \txcontract, \txfunction, \\ constructor\_o\_addr, constructor\_x, set\_x\}$, where:
\begin{itemize}
    \item $constructor\_o\_addr \hastype \addressT$
    \item $ constructor\_x$ and $set\_x \hastype \intT$
\end{itemize}
\end{itemize}

\Init is defined as follows:
\begin{align*}
 Init \defas \bigwedge_{a \in \Addresses} & \balanceSMT[a] \geq 0 \\
 \land \ & \constructed_{Bet} = \falsesmt \\
 \land \ & rate = 0 \\
 \land \ & oracle = \addressZero \\
 \land \ & player = \addressZero
 \end{align*}

Consider the procedure \txcode{join}. We have that (besides the implicit $s^0$ representing the payable modifier):
\begin{align*}
s^1 & \equiv \lastRevert^2 = 
\ITE(balance(C)^1 = 2\cdot\xavar \ \land \ player^i=\nullE, \ 	\falsesmt, \  \truesmt)
\\
s^2 & \equiv player^3 = \xavar
\end{align*}
%

\iftoggle{arxiv}{%
	\subsubsection{Encoding the property.}
}
{
	\subsection{Encoding the property}
}

Recall the \textbf{winnability} property $\phi_1$ from~\Cref{sec:logic:example}:
\begin{align*}
	\phi_1 = \;
	& (
	\contract{Bet}.\texttt{rate}>100 \,\land\, \contract{Bet}.\texttt{player}\neq \nullE
	)
	\rightarrow
	\big( 
	\exists a \hastype \addressT .\
	\exists f \hastype \methodT .\
	\exists x \hastype \calldataargsT .\
	\\
	& \txnext{{\txL{a}{\contract{Bet}}{f}{x}{0}}} \;\;
	\walE{a} = \oldL{\walE{a} + \walE{\contract{Bet}}}
	\big)
\end{align*}

\noindent
Let $\node$ be the node associated to $\phi_1$, 
let $\nodei$ be the node associated to: 
\[
\phi_1'\defas \walE{a} = \oldL{\walE{a} + \walE{\contract{Bet}}}
\]
and let $\edge$ be the connecting edge associated to $\txnext{{\txL{a}{\contract{Bet}}{f}{x}{0}}}$.

For simplicity, we assume that the contract has already been constructed, and we ignore the method \constructor.
\begin{align*}
	\Phi(\phi_1)
    = \;
	& (
	{rate^\node}>100 \,\land\, {player^\node}\neq \nullE
	)
	\rightarrow
	\\
	\Big( &
	\exists a \hastype \addressT .\
	\exists f \hastype \methodT .\
	\exists set\_x^x \hastype \intT .\
	 \\ &
	\exists \balanceSMT^\nodei  \hastype  \addressT \to \intT .\ \\ &
	\exists oracle^\nodei \hastype \addressT .\
	 \\ &
	 \exists rate^\nodei \hastype \intT .\ \\ &
	\exists \xavar^\edge \hastype \addressT  .\ \\ & 
	\exists \xnvar^\edge \hastype \intT  .\ \\ & 
	\exists \txfunction^\edge \hastype \methodT  .\ \\ & 
	\exists set\_x^\edge \hastype \intT .\
	\\
	& 
	\Trans
	\setenum{\bind{\StateVars}{\StateVars^\node}}
	\setenum{\bind{\TransVars}{\TransVars^\edge}}
	\setenum{\bind{\NextVars}{\StateVars^\nodei}}
	\\
	&   \land \ \xavar^\edge = a
	\\
	&   \land \ \txfunction^\edge = f
	\\
	&   \land \ set\_x^\edge = set\_x^x \\
	&   \land \ \xnvar^\edge = 0 
	\\
	& \land \balanceSMT[a]^\nodei = \balanceSMT[a]^\node + \balanceSMT[Bet]^\node
	\Big)
\end{align*}
}{}

\section{Evaluation}
\label{sec:evaluation}

\newcommand{\xmark}{\ding{55}}%

\newcommand{\statuslive}{\checkmark}
\newcommand{\statusnotlive}{\xmark}

\newcommand{\live}[1]{\ifempty{#1}{\statuslive}{\statuslive(#1)}}
\newcommand{\notlive}[1]{\statusnotlive(#1)}
\newcommand{\timeout}{T/O}
\newcommand{\tliveupto}{---}

We describe in this Section the implementation of our tool, the design of our benchmark, and the results of the evaluation.

\myparagraph{Implementation}
\ourtool takes as input a contract in our Solidity fragment and a set of \ourlogic properties, it type-checks and translates them to Lustre, and finally  model-checks with \kindtwo (\Cref{fig:overview}). 
The translation to Lustre follows the encoding described in~\Cref{sec:encoding}. 
Since \kindtwo does not support maps within properties, 
our translation  flattens map variables into integers (one per address).
By default, \kindtwo runs multiple engines in parallel, each employing a different model‑checking technique (e.g., bounded model checking~\cite{BMC}, k‑induction~\cite{kInd}, and IC3~\cite{IC3}). As back-end SMT solver, we selected \cvc.

\tikzset{every picture/.style={line width=0.75pt}} 

\begin{figure*}[t]
\centering
\resizebox{\textwidth}{!}{
\begin{tikzpicture}[x=0.75pt,y=0.75pt,yscale=-1,xscale=1]

\draw  [color={rgb, 255:red, 0; green, 0; blue, 0 }  ,draw opacity=1 ] (30,58) -- (75,58) -- (75,105.85) .. controls (46.88,105.85) and (52.5,123.11) .. (30,111.94) -- cycle ;
\draw    (87,83) .. controls (126.2,53.6) and (111.62,133.69) .. (148.67,108.67) ;
\draw [shift={(151,107)}, rotate = 143.13] [fill={rgb, 255:red, 0; green, 0; blue, 0 }  ][line width=0.08]  [draw opacity=0] (8.93,-4.29) -- (0,0) -- (8.93,4.29) -- cycle    ;
\draw   (157,106) .. controls (157,101.58) and (160.58,98) .. (165,98) -- (245,98) .. controls (249.42,98) and (253,101.58) .. (253,106) -- (253,130) .. controls (253,134.42) and (249.42,138) .. (245,138) -- (165,138) .. controls (160.58,138) and (157,134.42) .. (157,130) -- cycle ;
\draw    (265,117) .. controls (304.2,87.6) and (288.66,146.56) .. (325.67,120.69) ;
\draw [shift={(328,119)}, rotate = 143.13] [fill={rgb, 255:red, 0; green, 0; blue, 0 }  ][line width=0.08]  [draw opacity=0] (8.93,-4.29) -- (0,0) -- (8.93,4.29) -- cycle    ;
\draw   (354,95.5) .. controls (354,91.36) and (357.36,88) .. (361.5,88) .. controls (365.64,88) and (369,91.36) .. (369,95.5) .. controls (369,99.64) and (365.64,103) .. (361.5,103) .. controls (357.36,103) and (354,99.64) .. (354,95.5) -- cycle ;
\draw   (340,119.5) .. controls (340,115.36) and (343.36,112) .. (347.5,112) .. controls (351.64,112) and (355,115.36) .. (355,119.5) .. controls (355,123.64) and (351.64,127) .. (347.5,127) .. controls (343.36,127) and (340,123.64) .. (340,119.5) -- cycle ;
\draw   (370,125.5) .. controls (370,121.36) and (373.36,118) .. (377.5,118) .. controls (381.64,118) and (385,121.36) .. (385,125.5) .. controls (385,129.64) and (381.64,133) .. (377.5,133) .. controls (373.36,133) and (370,129.64) .. (370,125.5) -- cycle ;
\draw    (361.5,103) -- (350.02,110.38) ;
\draw [shift={(347.5,112)}, rotate = 327.26] [fill={rgb, 255:red, 0; green, 0; blue, 0 }  ][line width=0.08]  [draw opacity=0] (6.25,-3) -- (0,0) -- (6.25,3) -- cycle    ;
\draw    (361.5,103) -- (375.31,115.95) ;
\draw [shift={(377.5,118)}, rotate = 223.15] [fill={rgb, 255:red, 0; green, 0; blue, 0 }  ][line width=0.08]  [draw opacity=0] (6.25,-3) -- (0,0) -- (6.25,3) -- cycle    ;
\draw   (370,157.5) .. controls (370,153.36) and (373.36,150) .. (377.5,150) .. controls (381.64,150) and (385,153.36) .. (385,157.5) .. controls (385,161.64) and (381.64,165) .. (377.5,165) .. controls (373.36,165) and (370,161.64) .. (370,157.5) -- cycle ;
\draw    (377.5,133) -- (377.5,147) ;
\draw [shift={(377.5,150)}, rotate = 270] [fill={rgb, 255:red, 0; green, 0; blue, 0 }  ][line width=0.08]  [draw opacity=0] (6.25,-3) -- (0,0) -- (6.25,3) -- cycle    ;
\draw   (315,147.5) .. controls (315,143.36) and (318.36,140) .. (322.5,140) .. controls (326.64,140) and (330,143.36) .. (330,147.5) .. controls (330,151.64) and (326.64,155) .. (322.5,155) .. controls (318.36,155) and (315,151.64) .. (315,147.5) -- cycle ;
\draw   (347,147.5) .. controls (347,143.36) and (350.36,140) .. (354.5,140) .. controls (358.64,140) and (362,143.36) .. (362,147.5) .. controls (362,151.64) and (358.64,155) .. (354.5,155) .. controls (350.36,155) and (347,151.64) .. (347,147.5) -- cycle ;
\draw    (347.5,127) -- (325.16,138.62) ;
\draw [shift={(322.5,140)}, rotate = 332.53] [fill={rgb, 255:red, 0; green, 0; blue, 0 }  ][line width=0.08]  [draw opacity=0] (6.25,-3) -- (0,0) -- (6.25,3) -- cycle    ;
\draw    (347.5,127) -- (353.08,137.36) ;
\draw [shift={(354.5,140)}, rotate = 241.7] [fill={rgb, 255:red, 0; green, 0; blue, 0 }  ][line width=0.08]  [draw opacity=0] (6.25,-3) -- (0,0) -- (6.25,3) -- cycle    ;

\draw   (465,106) .. controls (465,101.58) and (468.58,98) .. (473,98) -- (553,98) .. controls (557.42,98) and (561,101.58) .. (561,106) -- (561,130) .. controls (561,134.42) and (557.42,138) .. (553,138) -- (473,138) .. controls (468.58,138) and (465,134.42) .. (465,130) -- cycle ;
\draw    (396,117) .. controls (435.2,87.6) and (419.66,146.56) .. (456.67,120.69) ;
\draw [shift={(459,119)}, rotate = 143.13] [fill={rgb, 255:red, 0; green, 0; blue, 0 }  ][line width=0.08]  [draw opacity=0] (8.93,-4.29) -- (0,0) -- (8.93,4.29) -- cycle    ;
\draw    (572,118) .. controls (611.2,88.6) and (611.02,148.52) .. (648.65,122.69) ;
\draw [shift={(651,121)}, rotate = 143.13] [fill={rgb, 255:red, 0; green, 0; blue, 0 }  ][line width=0.08]  [draw opacity=0] (8.93,-4.29) -- (0,0) -- (8.93,4.29) -- cycle    ;
\draw   (658,91) -- (703,91) -- (703,138.85) .. controls (674.88,138.85) and (680.5,156.1) .. (658,144.94) -- cycle ;
\draw   (848,118.5) -- (831.75,146.65) -- (799.25,146.65) -- (783,118.5) -- (799.25,90.35) -- (831.75,90.35) -- cycle ;
\draw    (858,101) .. controls (897.2,71.6) and (881.66,130.56) .. (918.67,104.69) ;
\draw [shift={(921,103)}, rotate = 143.13] [fill={rgb, 255:red, 0; green, 0; blue, 0 }  ][line width=0.08]  [draw opacity=0] (8.93,-4.29) -- (0,0) -- (8.93,4.29) -- cycle    ;
\draw    (858,137) .. controls (897.2,107.6) and (881.66,166.56) .. (918.67,140.69) ;
\draw [shift={(921,139)}, rotate = 143.13] [fill={rgb, 255:red, 0; green, 0; blue, 0 }  ][line width=0.08]  [draw opacity=0] (8.93,-4.29) -- (0,0) -- (8.93,4.29) -- cycle    ;
\draw    (858,120) .. controls (897.2,90.6) and (881.66,149.56) .. (918.67,123.69) ;
\draw [shift={(921,122)}, rotate = 143.13] [fill={rgb, 255:red, 0; green, 0; blue, 0 }  ][line width=0.08]  [draw opacity=0] (8.93,-4.29) -- (0,0) -- (8.93,4.29) -- cycle    ;
\draw  [fill={rgb, 255:red, 255; green, 255; blue, 255 }  ,fill opacity=1 ] (221.05,87) -- (272.5,87) -- (250.45,109) -- (199,109) -- cycle ;
\draw    (711,115) .. controls (750.2,85.6) and (734.66,144.56) .. (771.67,118.69) ;
\draw [shift={(774,117)}, rotate = 143.13] [fill={rgb, 255:red, 0; green, 0; blue, 0 }  ][line width=0.08]  [draw opacity=0] (8.93,-4.29) -- (0,0) -- (8.93,4.29) -- cycle    ;
\draw  [dash pattern={on 4.5pt off 4.5pt}] (133,41) -- (594,41) -- (594,175) -- (133,175) -- cycle ;
\draw   (36,141) .. controls (36,131.06) and (44.06,123) .. (54,123) .. controls (63.94,123) and (72,131.06) .. (72,141) .. controls (72,150.94) and (63.94,159) .. (54,159) .. controls (44.06,159) and (36,150.94) .. (36,141) -- cycle ;
\draw    (79,139) .. controls (118.2,109.6) and (111.3,156.07) .. (148.66,129.71) ;
\draw [shift={(151,128)}, rotate = 143.13] [fill={rgb, 255:red, 0; green, 0; blue, 0 }  ][line width=0.08]  [draw opacity=0] (8.93,-4.29) -- (0,0) -- (8.93,4.29) -- cycle    ;

\draw (53.09,42.5) node   [align=left] {Contract.sol};
\draw (205,118) node   [align=left] {Parser};
\draw (359,65) node   [align=left] {\begin{minipage}[lt]{57.35pt}\setlength\topsep{0pt}
\begin{center}
Abstract \\Syntax Tree
\end{center}

\end{minipage}};
\draw (513,118) node   [align=left] {Visitor};
\draw (681.09,75.5) node   [align=left] {Contract\&Properties.lus};
\draw (815.5,118.5) node   [align=left] {Kind2};
\draw (941,100) node   [align=left] {Valid};
\draw (944,123) node   [align=left] {Invalid};
\draw (941,143) node   [align=left] {T.O.};
\draw (235.75,98) node  [font=\footnotesize] [align=left] {ANTLR};
\draw (592,38) node [anchor=south east] [inner sep=0.75pt]   [align=left] {\begin{minipage}[lt]{36.75pt}\setlength\topsep{0pt}
\begin{center}
\ourtool
\end{center}

\end{minipage}};
\draw (54.09,168.5) node   [align=left] {Properties.hml};

\end{tikzpicture}
} 
\caption{Architecture of \ourtool.}
\label{fig:overview}
\end{figure*}

\myparagraph{Benchmark}
%
We evaluate our approach on a benchmark of representative use cases, covering different levels of complexity both in the contracts code and in the properties.
In the design of the properties, we focus in particular on those that are beyond the reach of current verification tools for Solidity.
%
For each use case we provide a main Solidity implementation and some mutations that introduce logic errors that may affect the validity of some the given properties.
For each verification task --- \ie, pair (property, mutation) ---
we provide the \emph{ground truth}, \ie a boolean telling whether the property holds or not for that mutation.
We summarize below the use cases and their main properties; their  \ourlogic formalization is in~\Cref{tab:benchmark}; their Solidity code is in~\Cref{sec:examples}.

\paragraph{Bank}

This is a wallet contract that allows users to deposit and withdraw tokens. 
The contract storage consists of a map that records the credits of each user.  
We study the following properties:
\begin{description}

\item[additivity] a user performing two consecutive deposits of, respectively, $n_1$ and $n_2$ tokens, obtains the same effect with a single deposit of $n_1+n_2$ tokens.

\item[reversibility] after a user deposits, the same user can perform a transaction that restores their balance to the one before the deposit.

\item[liquidity] any user with credits $n$ can always can fire a transaction that increases its balance of any $n_0 \leq n$; 

\item[frontrun deposit] front-running a user's deposit with someone else's transaction has no effect on the user's credits. 

\end{description}

Note that all these properties have a \emph{strategic} nature, in that they predicate about the existence of a sequence of transactions leading to some desired effect on the contract state.
To the best of our knowledge, no verification tool for Solidity can express general strategic properties such as the ones above
(see~\Cref{sec:related}).

\myparagraphB{Vault}
This contract allows the owner to deposit and withdraw tokens, with a mechanism to cancel pending withdrawal requests through a recovery key. 
Its behaviour is modelled as an automaton, \ie certain actions are enabled only in certain states and at certain times.
We consider the following properties:
\begin{description}

\item[two-step drainability] in idle state, some user can perform two transactions (possibly, at different blocks) to transfer the entire balance to some user.

\item[two-step non-inflation] in idle state, firing two transactions within a window of \code{wait\_time} blocks does not increase the balance of any user. 

\end{description}

As before, these are strategic properties involving universal and existential quantifiers (on the users and fired transactions), and are not supported by other verification tools besides ours.

\paragraph{Bet} 

This is a two-players bet on the price of a token, which is queried from an external oracle.
We consider the strategic properties introduced in~\Cref{sec:logic:example}.

\newcommand{\chmlscaling}{0.8}

\medskip
\newcommand{\chmlBankAdditivity}
{\scalebox{\chmlscaling}{$\begin{array}{l}
\forall a : \addressT.\ 
\forall c_1, c_2 : \intT.\ 
\exists v_{12},v_3:\intT.\ 
\exists r_1,r_2,r_3:\boolT.\;  
(c_1 \geq 0 \land c_2 \geq 0) 
\rightarrow
\\[1pt]
\;\;\Big(
\txnext{{\txL{a}{\contract{Bank}}{\code{deposit}}{}{c_1}}} \;
\Big(
r_1 = \lastrevL \; \land \;
\\[1pt]
\qquad
\txnext{{\txL{a}{\contract{Bank}}{\code{deposit}}{}{c_2}}} \; 
\big( v_{12} = \code{credits}[a] \land r_2 = \lastrevL \big)
\Big)
\\[1pt]
\land\;
\txnext{{\txL{a}{\contract{Bank}}{\code{deposit}}{}{(c_1+c_2)}}} \;\;
\big( v_{3} = \code{credits}[a] \land r_3 = \lastrevL \big)
\\[1pt]
\land\;
\big( r_1 \lor r_2 \lor (\neg r_3 \land v_{12} = v_3) \big)
\Big)
\end{array}$}}

\newcommand{\chmlBankReversibility}
{\scalebox{\chmlscaling}{$\begin{array}{l}
\forall a : \addressT . \ 
\forall c_1 : \intT . \;
\exists f: \methodT . \;
\exists xl: \calldataargsT . \;
\exists c_2 : \intT . \;\;
\\[1pt]
\quad
\txnext{{\txL{a}{\contract{Bank}}{\code{deposit}}{}{c_1}}} \;
\txnext{{\txL{a}{\contract{Bank}}{f}{xl}{c_2}}} \;
\big( 
\code{balance}[a] = \oldL{\oldL{\code{balance}[a]}}
\big)
\end{array}$}}

\newcommand{\chmlBankLiquid}
{\scalebox{\chmlscaling}{$\begin{array}{l}
\forall a : \addressT.\ 
\forall n : \intT.\ 
(n \geq 0 \land n \leq \code{credits}[a]) \rightarrow
\exists f : \methodT.\;
\exists xl : \calldataargsT.\;
\\[1pt]
\quad
\txnext{{\txL{a}{\contract{Bank}}{f}{xl}{0}}} \;
\big( 
\code{credits}[a] = \oldL{\code{credits}[a]} - n 
\land 
\code{balance}[a] = \oldL{\code{balance}[a]} + n \big)
\end{array}$}}

\newcommand{\chmlBankFrontrunDeposit}
{\scalebox{\chmlscaling}{$\begin{array}{l}
\forall a_1,a_2 : \addressT.\ 
a_1 \neq a_2 \rightarrow
\forall n_1,n_2 : \intT.\ 
\forall f : \methodT.\ 
\forall xl : \calldataargsT.\ 
\exists v_1, v_2 : \intT.\;
\\[1pt]
\quad
\Big(
\txnext{{\txL{a_1}{\contract{Bank}}{\code{deposit}}{}{n_1}}} \;
\big( v_1 = \code{credits}[a_1] \big)
\\[1pt]
\quad\land\;
\txnext{{\txL{a_2}{\contract{Bank}}{f}{xl}{n_2}}} \;
\txnext{{\txL{a_1}{\contract{Bank}}{\code{deposit}}{}{n_1}}} \;
\big( v_2 = \code{credits}[a_1] \big)
\\[1pt]
\quad\land\;
\big( v_1 = v_2 \big)
\Big)
\end{array}$}}

\newcommand{\chmlVaultA}
{\scalebox{\chmlscaling}{$\begin{array}{l}
\code{state}=0
\rightarrow
\exists a,b : \addressT . \;
\exists f_1,f_2 : \methodT . \;
\exists xl_1,xl_2: \calldataargsT . \;
\exists c_1,c_2 : \intT. 
\\[1pt]
\quad
\txnext{{\txL{a}{\contract{Vault}}{f_1}{xl_1}{c_1}}} \;
\txnext{{\txL{a}{\contract{Vault}}{f_2}{xl_2}{c_2}}} \;
\big(
\code{balance}[b] = \oldL{\oldL{\code{balance}[b] + \code{balance}}}
\big)
\end{array}$}}

\newcommand{\chmlVaultB}
{\scalebox{\chmlscaling}{$\begin{array}{l}
\code{state}=0 \rightarrow
\forall a,b,c : \addressT . \;
\forall f_1,f_2 : \methodT . \;
\forall xl_1,xl_2: \calldataargsT . \;
\forall c_1,c_2 : \intT. 
\\[1pt]
\quad
\txnext{{\txL{a}{\contract{Vault}}{f_1}{xl_1}{c_1}}} \;
\Big(
\code{block.number} \leq \oldL{\code{block.number}}+\code{wait\_time} 
\; \rightarrow
\\[1pt]
\quad\quad
\txnext{{\txL{b}{\contract{Vault}}{f_2}{xl_2}{c_2}}} \;
\big(
\code{balance}[c] = \oldL{\oldL{\code{balance}[c]}}
\big)
\Big)
\end{array}$}}

\newcommand{\chmlBetWinnability}
{\scalebox{\chmlscaling}{$\begin{array}{l}
(
\contract{Bet}.\texttt{rate}>100 \,\land\, \contract{Bet}.\texttt{player}\neq \nullE
)
\rightarrow
\big( 
\exists a \hastype \addressT .\
\exists f \hastype \methodT .\
\exists x \hastype \calldataargsT .\
\\[1pt]
\quad \txnext{{\txL{a}{\contract{Bet}}{f}{x}{0}}} \;\;
\walE{a} = \oldL{\walE{a} + \walE{\contract{Bet}}}
\big)
\end{array}$}}

\newcommand{\chmlBetLiquidity}
{\scalebox{\chmlscaling}{$\begin{array}{l}
\contract{Bet}.\texttt{player}\neq \nullE
\rightarrow  
\big(
\exists a_1,a_2 \hastype \addressT .\
\exists f_1,f_2 \hastype \methodT .\
\exists x_1,x_2 \hastype \calldataargsT .\
\\[1pt]
\quad 
\txnext{{\txL{a_1}{\contract{Bet}}{f_1}{x_1}{0}}} \;
\txnext{{\txL{a_2}{\contract{Bet}}{f_2}{x_2}{0}}} \;
\walE{\contract{Bet}} = 0
\big)
\end{array}$}}

\newcommand{\chmlBetFrontrunning}
{\scalebox{\chmlscaling}{$\begin{array}{l}
\forall a \hastype \addressT .\
\exists b \hastype \addressT.\
\exists f \hastype \methodT .\
\exists x \hastype \calldataargsT .\
\\
\quad
\txnext{{\txL{b}{\contract{Bet}}{f}{x}{0}}} \;
\txnext{{\txL{a}{\contract{Bet}}{\txcode{win}}{}{0}}} \;
\lastrevL
\end{array}$}}

\begin{table*}[t]
\centering
\small
\caption{Use cases and their properties.}
\label{tab:benchmark}
\resizebox{\linewidth}{!}{
\begin{tabular}{|c|c|l|}
\hline
\textbf{Use case} & \textbf{Property} & \multicolumn{1}{c|}{\textbf{\ourlogic specification}} \\
\hline
\multirow{12}{*}{Bank} 
& liquidity & \chmlBankLiquid \\
\cline{2-3}
& additivity & \chmlBankAdditivity
\\
\cline{2-3}
& reversibility & \chmlBankReversibility \\
\cline{2-3}
& \begin{tabular}{c} frontrun \\ deposit \end{tabular} & \chmlBankFrontrunDeposit  \\
\cline{2-3}
\hline
\multirow{5}{*}{Vault} 
& \begin{tabular}{c} two-step \\ drainability \end{tabular} & \chmlVaultA
\\
\cline{2-3}
& \begin{tabular}{c} two-step \\ non-inflation \end{tabular} & \chmlVaultB
\\
\hline
\multirow{3}{*}{Bet} 
& winnability & \chmlBetWinnability 
\\
\cline{2-3}
& liquidity & \chmlBetLiquidity
\\
\cline{2-3}
& frontrunning & \chmlBetFrontrunning
\\
\hline
\end{tabular}
}
\end{table*}


\myparagraph{Results}
\label{sec:evaluation:results}
%
%
\begin{table*}[t]
	\centering
    \scriptsize
	\caption{\ourtool results on our benchmark.}
	\label{tab:evaluation:results}
    \resizebox{0.9\linewidth}{!}{
	\begin{tabular}{|c|c|c|c|c|c|}
		\hline
		\textbf{Use case} & \textbf{Property} & \textbf{Mutation} & \textbf{Ground truth} & \textbf{Result} & \textbf{Time (s)} 
		\\
		\hline
		\multirow{16}{*}{Bank}
		& 
		\multirow{4}{*}{liquidity} & v1 & \statusValid & \valid{2} & 668.7
		\\
		\cline{3-6}
		&  & v2 & \statusValid & \valid{2} & 333.8
		\\
		\cline{3-6}
		&  & v3 & \statusInvalid & \invalid{1} & 0.3
		\\
		\cline{3-6}
		&  & v4 & \statusValid & \valid{2} & 512.1
		\\
		\cline{2-6}
		& 
		\multirow{4}{*}{additivity} & v1 & \statusValid & \valid{1} & 2.6
		\\
		\cline{3-6}
		&  & v2 & \statusInvalid & \invalid{0} & 0.1
		\\
		\cline{3-6}
		&  & v3 & \statusValid & \valid{1} & 17.3
		\\
		\cline{3-6}
		&  & v4 & \statusValid & \valid{1} & 3.4
		\\
		\cline{2-6}
		& 
		\multirow{4}{*}{reversibility} & v1 & \statusValid & \valid{2} & 21.6
		\\
		\cline{3-6}
		&  & v2 & \statusValid & \valid{2} & 26.7
		\\
		\cline{3-6}
		&  & v3 & \statusInvalid & \invalid{0} & 0.2
		\\
		\cline{3-6}
		&  & v4 & \statusValid & \valid{2} & 31.9
		\\
		\cline{2-6}
		& 
		\multirow{4}{*}{frontrun deposit} & v1 & \statusValid & \valid{1} & 22.1
		\\
		\cline{3-6}
		&  & v2 & \statusValid & \valid{1} & 26.9
		\\
		\cline{3-6}
		&  & v3 & \statusValid & \valid{1} & 307.8
		\\
		\cline{3-6}
		&  & v4 & \statusInvalid & \invalid{0} & 0.3
		\\
		\hline
		\multirow{6}{*}{Vault}
		& 
		\multirow{3}{*}{two-step drainability} & v1 & \statusValid & \valid{2} & 33.9
		\\
		\cline{3-6}
		&  & v2 & \statusInvalid & \invalid{0} & 0.2
		\\
		\cline{3-6}
		&  & v3 & \statusValid & \valid{2} & 34.4
		\\
		\cline{2-6}
		& 
		\multirow{3}{*}{two-step non-inflation} & v1 & \statusValid & \valid{1} & 0.4
		\\
		\cline{3-6}
		&  & v2 & \statusValid & \valid{1} & 0.4
		\\
		\cline{3-6}
		&  & v3 & \statusInvalid & \invalid{0} & 0.1
		\\
		\hline
		\multirow{9}{*}{Bet}
		& 
		\multirow{3}{*}{winnability} & v1 & \statusValid & \valid{2} & 84.2
		\\
		\cline{3-6}
		&  & v2 & \statusInvalid & \invalid{1} & 0.3
		\\
		\cline{3-6}
		&  & v3 & \statusInvalid & \invalid{1} & 0.3
		\\
		\cline{2-6}
		& 
		\multirow{3}{*}{liquidity} & v1 & \statusValid & \valid{5} & 90.6
		\\
		\cline{3-6}
		&  & v2 & \statusInvalid & \invalid{1} & 0.4
		\\
		\cline{3-6}
		&  & v3 & \statusValid & \valid{5} & 97.3
		\\
		\cline{2-6}
		& 
		\multirow{3}{*}{frontrunning} & v1 & \statusValid & \valid{2} & 166.1
		\\
		\cline{3-6}
		&  & v2 & \statusValid & \valid{1} & 0.3
		\\
		\cline{3-6}
		&  & v3 & \statusInvalid & \invalid{1} & 0.3
		\\
		\hline
	\end{tabular}
    }
\end{table*}

%
%
%
We run \ourtool on each verification task on a 3GHz 64-bit Intel Xeon Gold 6136 CPU and GNU/Linux OS (x86\_64-linux) with 64 GB of RAM,
with \kindtwo (v2.3.0-ge8216dd)
and \cvc (v1.1.3-dev.152.701cd63ef) as backend SMT solver.
The run-time limit for each veriﬁcation task is \timelimit s of CPU time.

A subset of the results is shown in~\Cref{tab:evaluation:results}. 
We mark each verification task as: ``\invalid{N}'', if the solver finds a trace that violates the property (with $N$ being the length of the shortest trace leading to a violation), and ``\valid{N}'', if it proves that the property holds in all possible states (with $N$ being the length of the $k$-induction base case).
%
%

\myparagraph{Discussion} 
The results show that \ourtool always returns an answer consistent with the ground truth, and that it is effective both at proving the validity of the property as well as finding violations.
When it finds a counterexample, it does so extremely quickly (always in less than 1 second), and it returns as witness a sequence of transactions that can be replayed in the actual Ethereum, leading to a state from which the desired outcome is unreachable 
(an example can be found \href{\exCex}{here}). 
Proving that a property is valid can be more time-consuming, as it requires performing advanced deductive reasoning based on inductive techniques.
When a property is proven to be valid, the user has the mathematical guarantee that it cannot be violated in \ourtool semantics. 
Given that \ourtool abstracts some semantical aspects of Solidity (\eg reentrancy), 
it is not guaranteed that, moving to the actual Ethereum blockchain,  the property will preserved ``as-is''; 
however, the output of \ourtool guarantees that no conceptual error has been made in the business logic of the contract.



\section{Related work}
\label{sec:related}



Analysis tools for smart contract can be roughly partitioned in two classes: \emph{vulnerability detection} tools, which target pre-defined classes of bugs (\eg, reentrancy, overflows, \etc) and \emph{verification} tools, that instead can statically determine whether user-defined properties are satisfied in any contract executions.

While vulnerability detection tools are not designed to capture violations of user-defined temporal properties, some of them can detect restricted instances of liquidity and front-running vulnerabilities through ad-hoc analyses tailored to specific patterns.
In that setting, liquidity vulnerabilities are often referred to as ``Locked Ether'', which roughly means that there is some contract state where it is impossible for anyone to withdraw Ether from the contract.
The tools capable of detecting this vulnerability include   
Slither~\cite{slither},
SmartCheck~\cite{smartcheck},
Maian~\cite{maian},
Securify2~\cite{Tsankov18ccs},
ConFuzzius~\cite{confuzzius}
and sFuzz~\cite{sfuzz}.
Each tool, however, has its own interpretation of the property, making a comparison difficult~\cite{Sendner24sp}.
A similar situation occurs for front-running, often referred to as \emph{transaction order dependence} in vulnerability classifications~\cite{SWC114}.
A few tools can flag patterns in which the outcome of a transaction depends on shared state that can be modified by a competing transaction.
For instance, 
Oyente~\cite{Luu16ccs}, 
Securify~\cite{Tsankov18ccs}, 
EthRacer~\cite{ethracer} 
and SailFish~\cite{sailfish}
introduce specific patterns to detect when Ether transfers (including the amount and receiver) are affected by transaction ordering.

There are two key differences between these tools and ours:
\begin{itemize}

\item While vulnerability detection tools can provide smart contract developers with a useful feedback, spotting parts of the code that may contain potential vulnerabilities, they cannot determine that a contract enjoys some desirable property across all possible executions.
By contrast, \ourtool is a formal verification tool, and as such it can establish that a given property holds universally, or up to a given bound on the number of transactions.

\item Vulnerability detection tools typically lack flexible mechanisms for specifying the properties to be analysed. 
In most cases, they target predefined, hard-coded properties, offering little or no programmability. 
By contrast, \ourtool allows developers to verify arbitrary \ourlogic specifications.
As shown in~\Cref{sec:evaluation}, this flexibility is crucial to establish relevant properties of smart contracts, such as general  liquidity properties predicating on \emph{who} can withdraw, \emph{how much}, and under \emph{which conditions}. 

\end{itemize}


Moving to verification tools, many support only safety properties, including \eg of SolCMC~\cite{AltS22cav}, Zeus~\cite{Kalra18ndss},
solc-verify~\cite{Hajdu19vstte}, 
SmartACE~\cite{Wesley22vmcai} and
VerX~\cite{Permenev20sp}.
We discuss below tools that handle a broader class of properties.


The \emph{Certora Prover}~\cite{certora} is one of the leading formal verification tools for Solidity.
The tool takes as input a Solidity contract and a specification in its domain-specific language (CVL), which defines constraints on the execution of a contract and assertions that must be true in all states satisfying the given constraints.
Certora compiles the Solidity contract and the CVL spec into a logical formula, and relies on off-the-shelf SMT solvers to determine if the spec is satisfied in all contract states.
Compared to \ourtool, the Certora Prover is an industrial-strength tool that supports the full Solidity language, while ours is a prototype that targets a representative fragment.
This design choice impacts the way certain properties are formalized. 
For example, in \ourlogic one can directly specify that a transaction transfers a given amount of ETH to an account by constraining the account balance before and after the transaction.
In other words, \ourtool assumes that ETH transfers neither fail nor propagate the received tokens.
Expressing the same property in CVL is quite problematic, since in full Solidity ETH transfers are rendered as external calls, which may trigger further calls that propagate additional transfers or revert the execution~\cite{BFMPS24fmbc}.
Another key difference between CVL and \ourlogic concerns the treatment of quantification. 
In CVL, all free variables in a rule are implicitly universally quantified, while \ourlogic supports arbitrary nesting of universal and existential quantifiers. 
This flexibility is essential for expressing strategic properties of contracts.
For instance, for liquidity we want to specify that \emph{for all} users there \emph{exists} some transaction that the user can execute to achieve a desired effect on their balance.
Mixing universal and existential quantifiers is also needed to express properties such as reversibility (cf.~\Cref{tab:benchmark}).
%
Regarding verification, \ourtool and the Certora Prover adopt different approaches to reporting property violations.
\ourtool provided a concrete witness in the form of a sequence of transactions that, starting from an initial blockchain state, leas to a state in which the specified invariant is false.
This implies that violations detected by \ourtool can be translated into executable proofs-of-concept (up-to the abstractions introduced by the considered Solidity fragment).
By contrast, when the Certora Prover reports a violation, it witnesses it through a blockchain state that is not guaranteed to be reachable.  
While this approach speeds-up verification, it has the drawback that negative verification reports cannot be directly translated into concrete exploits; moreover, in some cases, a property reported as violated may in fact hold in all reachable executions.


\emph{SmartPulse}~\cite{Stephens21sp} verifies Solidity contracts against specifications written in a language based on Linear Temporal Logic (LTL).
While this enables expressing liveness properties, it does not capture liquidity and frontrunning, which lie beyond LTL.
Thus, SmartPulse and \ourtool have uncomparable expressiveness.
Moreover, practical use of liveness properties requires a \emph{fairness assumption}, \ie another LTL formula specifying the traces in which users perform the actions needed to reach a desired state.
In SmartPulse, this formula must be anticipated and encoded by the designer.
Instead, \ourtool can automatically infer the transaction parameters that produce the desired state change.


\emph{VeriSolid}~\cite{Nelaturu23tdsc} takes as input a Solidity contract and its properties expressed in Computation Tree Logic (CTL), transforms the contract into an Abstract State Machine (ASM), and verifies the properties against the ASM using tools in the BIP toolchain, such as the {nuXmv} symbolic model checker~\cite{Bliudze15atva}.
The liveness properties specified in~\cite{Nelaturu23tdsc} are not accompanied by fairness assumptions (unlike SmartPulse), but in principle this seems doable without reworking the verification techniques
(we note that a more recent extension of VeriSolid includes fairness assumptions~\cite{Chahoki23overlay}).
Liveness properties expressed in CTL, however, cannot encompass the strategic properties addressed by \ourtool. 
In particular, \ourlogic allows one to mix universal and existential quantification, which seems beyond the scope of the version of CTL supported by VeriSolid.


\emph{Solvent}~\cite{solvent-ifm} is a verification tool targeted to liquidity properties. 
It translates a fragment of Solidity similar to the one considered here into SMT constraints~\cite{SMTHandbook}.
The set of properties expressible by Solvent is a subset of \ourlogic.
In particular, Solvent supports properties with a specific pattern: a universal quantification, followed by an existential quantification over a transactions;
however, while this pattern is sufficient to characterise typical liquidity properties
(\eg, ``every user can fire a transaction that produces a certain effect''), it does not cover more complex properties that require arbitrary combinations of universal and existential quantifiers, such as most of those in~\Cref{tab:benchmark}.
Another difference between Solvent and \ourtool concerns the techniques used.
Indeed, Solvent 
relies on bounded model checking (BMC) to return a counterexample (if the property is violated), or show that, up to a certain number of transactions, the property holds. 
Except that in simple cases, in which naive predicate abstraction suffices, Solvent is not able to prove that a property holds for \emph{all} reachable states.
On the contrary, \ourtool leverages  all  the advanced model checking techniques implemented in \kindtwo, enabling the tool to prove validity also for complex instances, besides benefiting from well-rounded heuristics.



Besides \emph{formal} verification tools, recent work has explored the use of LLMs to assess the validity of arbitrary contract-specific properties written in natural language~\cite{BLPfc26}. 
This approach allows users to investigate ---  potentially --- \emph{any} class of properties, including those not expressible in \ourlogic (\eg, properties mentioning arbitrarily long sequences of transactions or assumptions on contract-to-contract calls).
However, the approach has two significant weaknesses: natural language properties are intrinsically ambiguous, and there is no
guarantee that the answers provided by the LLMs are correct.
In contrast, \ourtool  follows a principled formal method approach, which guarantees the correctness of its  answers.

\paragraph{\ourlogic \emph{vs.} other  logics.}
	%
%
%
%
%
%
%
%
%
%
%
Several modal logics allow predicates to range not only over \emph{states} but also  over \emph{actions}. 
Among these, one of the most influential is Hennessy-Milner logic (HML)~\cite{HennessyMilner85jacm}, a dynamic logic interpreted over LTSs.
HML introduces the modal constructs $\txnext{\Act}\,\phi$ and $\alltxnext{\Act}\phi$ --- where $\Act$ is a set of actions and $\phi$ a propositional formula --- to express the conditions ``\emph{there exists an action $a \in \Act$ \normalfont{(resp., ``\emph{for all actions}'')}, after which $\phi$ holds}''.
Since in HML only properties of finite depth can be described,
an extension of HML with recursion~\cite{HMLrecursion} (equivalent to the modal $\mu$-calculus~\cite{ReactiveSystemsBook}) has been proposed, enabling the expression of a wide range of temporal properties, including safety and liveness. 
Another temporal logic that allows to predicate over actions is Action CTL* (sometimes denoted ACTL*)~\cite{ActionCTL}, whose expressive power lies between that of HML and HML with recursion. 
Strategic logics such as ATL~\cite{AlurHK02} extend temporal logics to reason about \emph{strategies} of agents in multi-agent systems. 

There are several differences between \ourlogic and the previous logics. 
Compared to HML, the verification problem over \ourlogic asks whether the property holds \emph{in every reachable state} (hence, for example, safety properties are expressible).
Compared to HML with recursion and to Action CTL*, \ourlogic is more restrictive,
 as it does not allow branching over arbitrarily long paths nor the use of the \emph{finally} operator (hence liveness is not expressible). 
 %
 %
 Using the terminology of Action CTL*, the verification of properties in \ourlogic can be thought as corresponding to the verification of properties of the form 
 ``{$\mathrm{A\,G\,}\phi$}'', 
 where $\phi$ can contain arbitrary combinations of 
 ``$\mathrm{E_\alpha\,X\,}$'' 
 and 
 ``$\mathrm{A_\alpha\,X\,}$''
 modal operators (the $\alpha$ subscript denotes predication over actions).
Compared to strategic logics such as ATL and its extensions (\eg, WATL~\cite{FerrandoKM26}), \ourlogic can express ``strategic properties'' of the form ``\emph{in every reachable state, a user (or a group of users) can always reach a desired state in a finite amount of steps}'' (\eg liquidity),  
but it cannot express properties that predicates over  \emph{users strategies} intended as maps from (sequences of) states to actions.
Additional differences with the previous logics are that \ourlogic extends propositional logic to first-order (FO) logic, and that it features a \emph{past} operator.
While temporal logics extended to FO (\eg, FOLTL~\cite{FOLTL} and FOCTL~\cite{FOCTL}) and with past operators~\cite{pastTempLogic}  have already been studied,
 \ourlogic features a combination of action-based operators, past operator, and first-order logic 
(plus domain-specific constructs), 
which is
specifically
tailored to capture relevant properties of smart contracts while remaining practically addressable by automatic formal verification tools. 
To the best of our knowledge, no existing logic is equivalent to \ourlogic.

\section{Conclusions and Future work}
\label{sec:conclusions}

We proposed \ourlogic, a novel logic for expressing complex temporal properties of smart contracts, and \ourtool, a tool for verifying \ourlogic properties of smart contracts.   
Our experiments show that \ourtool is capable of verifying or refuting properties that are beyond the reach of existing tools.

Several directions remain for future work. 
In particular, the current version of \ourtool is limited to single-contract properties, whereas many relevant properties involve interactions among multiple contracts: for instance, a user must extract tokens from a contract to enable an action on another.  
A lightweight approach to handle this in \ourtool would consist in embedding the functions of multiple contracts into a single one.
However, doing this requires some care: \eg, the balances of the embedded contracts must be kept distinct, and the visibility of contract functions must be taken into account. 
Another research line is extending the Solidity fragment supported by \ourtool in order to narrow the gap with the full Solidity language.
Along this line, significant extensions would be the treatment of contract-to-contract calls (including those triggered by simple ETH transfers), loops, and the gas mechanism.

Another direction for future work is the automatic transformation of the counterexamples produced by \ourtool{} into executable proofs-of-concept (PoCs). In the context of security analysis, PoCs play a crucial role by demonstrating that a reported vulnerability is not merely theoretical but can be concretely exploited in practice.  Constructing PoCs typically requires significant manual effort and expertise, as one must derive a sequence of transactions that reproduces the exploit scenario in a realistic execution environment. 
Recent research has investigated the automatic generation of PoCs for  smart contracts, leveraging techniques such as symbolic reasoning over execution traces, and large language models capable of synthesizing exploit scripts from vulnerability descriptions or execution traces~\cite{Prompt2Pwn,Gervais26fc}.
\enriconote{tutta questa roba non è banale nel contesto delle nostre proprietà, le PoC vanno bene fondamentalmente per simil-safety, se vai oltre ti servono proof...}
Integrating similar techniques into \ourtool would enable the automatic synthesis of concrete exploit transactions directly from model-checking counterexamples, producing PoCs that can be executed on a local blockchain or test network. Such functionality would not only facilitate debugging and validation of detected violations, but would also provide actionable artifacts for developers and auditors, bridging the gap between formal verification results and practical security analysis.
In our setting, this appears particularly promising because the counterexamples generated by the model checker already encode sequences of contract interactions that violate the desired temporal property, providing a natural starting point for synthesizing executable exploit transactions.

\paragraph*{Acknowledgments}
Massimo Bartoletti and Enrico Lipparini have been partially supported by the projects PRIN 2022 DeLiCE (F53D23009130001) and SERICS (PE00000014) under the MUR National Recovery and Resilience Plan funded by the European Union --- NextGenerationEU. 

\bibliographystyle{splncs04}
\bibliography{main}

\newpage

\iftoggle{arxiv}{}
{%
\appendix

}

\end{document}